\documentclass[aps,pra, superscriptaddress]{revtex4-2}

\usepackage{graphicx}
\usepackage{amsmath,amssymb}
\usepackage{hyperref}
\usepackage{lineno}
\usepackage{physics}
\usepackage{bm}
\usepackage{siunitx}
\usepackage{xcolor}
\newcommand{\cint}{C_{\mathrm{int}}}
\newcommand{\cintdirect}{C_{\mathrm{int}}}
\newcommand{\mincint}{C_{\mathrm{int}}^{\mathrm{min}}}

\newcommand{\kint}{\kappa_{\mathrm{int}}}
\newcommand{\tcrit}{t_{\mathrm{crit}}}
\newcommand{\dcrit}{D_{\mathrm{crit}}}
\newcommand{\scrit}{s_{\mathrm{crit}}}

\newcommand{\myvector}[1]{\Vec{#1}}
\newcommand{\mymatrix}[1]{\hat{#1}}

\newcommand{\naturalnumbers}{\mathbb{N}_0}
\newcommand{\integers}{\mathbb{Z}}
\newcommand{\retroreflectivetargetmode}{\epsilon_{\mathrm{T}}}
\newcommand{\retroreflectiveeigenmode}{\epsilon_{\mathrm{E}}}

\newcommand{\lint}{\mathcal{L}_{\mathrm{int}}}
\newcommand{\lclip}{\mathcal{L}_{\mathrm{clip}}}
\newcommand{\lnotclip}{\mathcal{L}_{\mathrm{abs}}}

\newcommand{\inverseeffectivearea}{A^{-1}_{\mathrm{eff}}}
\newcommand{\inverseeffectiveareaemitter}{\inverseeffectivearea(\bm{r_e})}
\newcommand{\lintmatrix}{\mymatrix{\mathcal{L}}_{\mathrm{int}}}
\newcommand{\inverseareamatrix}{\mymatrix{A}^{-1}_{\mathrm{eff}}}
\newcommand{\pext}{P_{\mathrm{ext}}}
\newcommand{\shapingscale}{(1/\lambda_s)_{\mathrm{crit}}}
\newcommand{\inverseeffectivelength}{L^{-1}_{\mathrm{eff}}}
\newcommand{\inverseeffectivelengthemitter}{\inverseeffectivelength(\bm{r_e})}
\newcommand{\inverseeffectivelengthy}{L^{-1}_{\mathrm{y, eff}}}
\newcommand{\basissizespherical}{N_{\mathrm{sph.}}}
\newcommand{\samplesspherical}{s_{\mathrm{sph.}}}
\newcommand{\basissize}{N}
\newcommand{\samples}{s}
\newcommand{\basissizeopt}{N_{\mathrm{opt.}}}
\newcommand{\inverselengthmatrix}{\mymatrix{L}^{-1}_{\mathrm{eff}}}
\newcommand{\transversecoordinate}{\bm{r}_{\perp}}

\begin{document}
\title{Retroreflective surface optimisation for optical cavities with custom mirror profiles}

\author{William J. Hughes}
\email{w.j.hughes@soton.ac.uk}
\author{Peter Horak}
\affiliation{Optoelectronics Research Centre, University of Southampton, SO17 1BJ, UK}

\date{\today}

\begin{abstract} 
Coupling an emitter to a Fabry-P\'{e}rot optical cavity can provide a coherent and strong light-matter interface whose performance in a variety of applications depends critically on the emitter-photon coupling strength. Altering the typically spherical profiles of the cavity mirrors can improve this coupling strength, but directly optimising the mirror shape is numerically challenging as the multidimensional parameter space features many local optima. Here, we develop a two-step method to optimise mirror surface profiles while avoiding these issues. First, we optimise the target cavity eigenmode for the chosen application directly, and second, we construct the mirror surfaces to retroreflect this optimised target mode at both ends of the cavity. We apply our procedure to different emitter-cavity coupling scenarios. We show that mirror shaping can increase the cooperativity of coupling to a central emitter by a factor of approximately 3 across a wide range of geometries, and that, for coupling two or more emitters to a single cavity mode, the improvement factors can far exceed an order of magnitude.

\end{abstract}

\maketitle

\section{Introduction}
Fabry P\'{e}rot optical cavities can offer greatly-enhanced, and even near-deterministic interfaces between quantum emitters and single photons~\cite{Keller:04, McKeever:04, Bochmann:10, Grinkemeyer:25} while preserving high optical access~\cite{Hartung:24}. These improvements promise significant performance increases in many applications of quantum technology, such as quantum networking~\cite{Ritter:12, Reiserer:15, Krutyanskiy:23b}, modular quantum computation~\cite{Monroe:14, Li:24}, and quantum-enhanced sensing~\cite{Petrak:14}. These cavities have typically been built with spherical mirrors~\cite{Law:97, Krutyanskiy:19}, yielding Gaussian cavity modes with well-known properties~\cite{Siegman:86, Yariv:91}. However, proposals have suggested that non-spherical mirrors could confer advantageous features to the cavity modes~\cite{Ferdous:14, Podoliak:17, Karpov:22_2, Karpov:23}. As the field pivots towards microcavities in an effort to reduce footprint, improve robustness and scalability, and to reduce interaction timescales~\cite{Steiner:13, Saavedra:21, Pfeifer:22, Teller:23, Zifkin:24}, alternative methods to traditional superpolishing are required to create the highly curved mirrors~\cite{Steiner:13, Jin:22}. This makes mirror shaping an increasingly compelling prospect, as micromirror fabrication techniques like focused ion beam milling~\cite{Dolan:10} and, to a lesser extent, laser ablation~\cite{Steinmetz:06, Hunger:10} offer control over the surface topography~\cite{Trichet:15, Walker:21, Maier:25, Ott:16, Gao:25}.

However, the optimisation of a cavity mirror surface is numerically difficult because the relation between the mirror topography and the structure of the resonant modes is not straightforward (outside of perturbation assumptions~\cite{vanExter:22} and specific geometric regimes~\cite{Nyman:14}). This means that any reasonably flexible parameterisation of the surface features a vast parameter space which typically contains many local optima. It is therefore difficult to design the optimum surface for a given task through direct optimisation of the topography without working in a limited parameter space~\cite{Karpov:22_2} that may hide the possibility for performance improvements.

In this manuscript, we present a two-stage procedure to optimise a cavity for a user-specified goal. The first step is to optimise the cavity mode, instead of the mirror surface, to take advantage of the (typically) simpler optimisation landscape. The second step is to construct the cavity surface to retroreflect this target mode. We show that this approach finds cavities whose performance exceeds any cavity with spherical mirrors, with the potential improvements particularly stark for cavities that are designed to couple to multiple emitters simultaneously.

\section{Mode parameterisation and optimisation}
\label{Sec: mode expansion theory}
Our method will optimise the optical field inside Fabry-P\'{e}rot cavities, and we therefore need to describe a general intracavity field. We assume that this intracavity field is monochromatic and not highly divergent from the cavity axis, which is generally a good assumption for Fabry-P\'{e}rot cavities that are typically longer than they are wide. Under these assumptions, the intracavity field can be described using the paraxial equation. Formally, the intracavity electric field is a superposition of forward and backward running waves, $\bm{E}^{\pm}(\bm{r},t)$ at spatial position $\bm{r}=(x,y,z)$ and time $t$, which are specified through a scalar field $\epsilon^{\pm}(\bm{r})$~\cite{Barre:17} that satisfies
\begin{align}
    \bm{E}^{\pm}(\bm{r},t) & = \bm{{\epsilon}}^{(u)} \epsilon^{\pm}(\bm{r})\exp(\mp ikz)\exp(i\omega t), \label{eq: paraxial field definition} \\
    \pdv{z}\epsilon^{\pm}(\bm{r}) & = \mp\frac{i}{2k}\left(\pdv[2]{x} + \pdv[2]{y}\right)\epsilon^{\pm}(\bm{r}). \label{eq: paraxial equation}
\end{align}
with $\omega$ the angular frequency of the field, $k=\omega/c$ the wavevector, $\bm{{\epsilon}}^{(u)}$ the fixed, unit-magnitude polarisation of the field, which must be 
perpendicular to the $z$-axis, and $\pm$ indicating propagation towards or away from positive $z$ respectively.

The paraxial equation (Eq.~(\ref{eq: paraxial equation})) is satisfied by Gaussian beam solutions, which are particularly convenient for Fabry-P\'{e}rot cavities because they are transversely localised and satisfy the reflection boundary conditions imposed by spherical mirrors. We will be largely concerned with radially symmetric mirrors, and will therefore use the Laguerre-Gauss modes, which may be written

\begin{equation}
\begin{aligned}
\epsilon^{LG, \pm}_{l,m}\left(\bm{r}\right)  = &  \sqrt{\frac{2l!}{\pi\left(l + \abs{m}\right)!}}  \frac{1}{w(z)} \left(\frac{\sqrt{2}r}{w(z)}\right)^\abs{m} L^{\abs{m}}_l \left(\frac{2r^2}{w^2(z)}\right) \\
& \exp\left\{-\frac{r^2}{w^2(z)} -i \left(m\phi \pm \frac{kr^2}{2R_u(z)} \mp \left(2l + \abs{m} + 1\right)\psi_G \right)\right\}, 
\end{aligned}
\label{eq: Hermite Gauss mode}
\end{equation}
where $r$, $\phi$, and $z$ are the conventional coordinates of a cylindrical polar system, with
\begin{equation}
    \begin{aligned}
        w(z)  = w_0 \sqrt{1+\left(\frac{z}{z_0}\right)^2}, \quad & z_0 = \frac{\pi w_0^2}{\lambda}, \\
R_u(z)   = z\left(1+\left(\frac{z_0}{z}\right)^2\right), \quad &
\Phi_G(z)  =  \arctan\left(\frac{z}{z_0}\right),
    \end{aligned}
\end{equation}
where $\lambda=2\pi/k$ is the wavelength, $L^\alpha_n$ are the associated Laguerre polynomials, and $z_0$ is the Rayleigh range of the beam. The coordinate origin has been fixed to lie at the centre of the smallest waist of the beam. The Laguerre-Gauss modes are indexed by radial index $l\in\naturalnumbers$ and azimuthal index $m\in\integers$.

We will also optimise cavities in one transverse dimension (i.e. suppressing functional dependence in, and derivatives of $y$). For these cases, the paraxial equation can be solved with Hermite-Gauss modes

\begin{equation}
    \epsilon^{HG, \pm}_{n}\left(\bm{r}\right) =  \sqrt{\sqrt{\frac{2}{\pi}}\frac{1}{w(z)2^{n}n!}}H_{n}\left(\frac{\sqrt{2}x}{w(z)}\right)\exp\left\{-\frac{x^2}{w^2(z)} -i\left(\pm \frac{kx^2}{2R_u(z)} \mp \left(n+\frac{1}{2}\right)\Phi_G(z)\right)\right\},
\end{equation}
where $H_{n}$ are the physicist's Hermite polynomials and $n \in \naturalnumbers$ is the transverse index of the mode. Both Laguerre-Gauss and Hermite Gauss states form complete basis sets (Laguerre-Gauss over a 2D transverse plane, Hermite-Gauss modes over a single transverse dimension)~\cite{Nienhuis:93}. This means that any field satisfying the paraxial equation can be written as a linear superposition of the basis modes in the appropriate domain,
\begin{equation}
    \epsilon^{\pm}(\bm{r}) = \sum_{s=0}^S C_s \epsilon^{\Lambda, \pm}_{s}(\bm{r}),
    \label{eq: Laguerre Gauss expansion}
\end{equation}
provided the basis size $S+1$ is sufficiently large, where $\Lambda$ is either $LG$ or $HG$, the mode indices $(l,m)$ or $n$ are mapped to a single index $s$, and $C_s$ is the complex coefficient of the basis state $\epsilon^{\Lambda,\pm}_s$. We can thus identify any solution to the paraxial equation with a complex vector $\myvector{C}$
\begin{equation}
    \myvector{C} = \{C_0, C_1, ..., C_S\}^T.
\end{equation}
Our method will optimise the cavity mode for a given task by optimising the coefficients of this vector.

The process by which the coefficients of $\myvector{C}$ are optimised depends upon the metric of performance that is being optimised. For the general case where only the performance metric (but not the gradient) can be calculated for a given set of coefficients, one may use local or global optimisation. For a substantial number of interesting cases, the performance metric may be written as the product and quotient of expectation values of Hermitian matrices. In this case, the gradient can be evaluated directly (see App.~\ref{app: gradient based retroreflective optimisation}), which allows for faster local optimisation.

 \section{Obtaining mirror topography from retroreflection}
\label{sec: retroreflective surface optimisation}
In order to correspond an intracavity vector field $\myvector{C}$ to a mirror surface, we build upon the technique of Karpov et al.~\cite{Karpov:22_1} for designing surfaces to enhance emitter-cavity coupling strength. In that technique, a target eigenmode. henceforth known as $\retroreflectivetargetmode$, with strong coupling to the emitter was chosen, and the corresponding mirror \textit{constructed} by matching its surface to the equiphase surface contour of the target mode. If these bespoke mirrors were of infinite extent, they would perfectly retroreflect the chosen mode, which would remain unchanged (up to a phase) upon a round trip of the cavity, making $\retroreflectivetargetmode$ an eigenmode of the cavity. Here, instead of starting with an ansatz for $\retroreflectivetargetmode$ designed to improve coupling to a central emitter, we optimise the $\retroreflectivetargetmode$ to maximise a metric of performance that is chosen according to the application at hand.

To construct the mirror surface, we need to find the equiphase contour of the target electric field. The phase of the electric field is composed of the phase of the paraxial mode $\retroreflectivetargetmode$, and the propagating wave phase $\exp(\mp ikz)$ (see Eq.~(\ref{eq: paraxial field definition})). For each coordinate $\bm{r}=(\transversecoordinate, z_{m_{\pm}})$ on the plane for the positive (negative) $z$ mirror, where $\transversecoordinate$ is some parameterisation of the transverse coordinate, the incoming paraxial mode $\epsilon^+_{T}$ has a phase, whose differences across the mirror profile should be offset by the running wave phase incident on the varying surface protrusion $f_{\pm}(\transversecoordinate)$ which, by convention, points towards the centre of the cavity. Mathematically,
\begin{equation}
    kf_{\pm}(\transversecoordinate)\mod2\pi = \mp\mathrm{arg}\left[\epsilon^{+}_{T}(r,z_{m_{\pm}})\right].
    \label{eq: surface reconstruction}
\end{equation}
Strictly speaking, $f_{\pm}(\transversecoordinate)$ is only an equiphase contour on the assumption that the phase of $\retroreflectivetargetmode$ changes only negligibly as $z$ changes between the nominal mirror plane and the protruded surface. We do not correct this assumption because the discrepancy generally has little impact, and because the mode mixing method, which we use to calculate the cavity eigenmodes, also makes this assumption~\cite{Hissink:22}.

There is, however, a complication with the logic of retroreflective optimisation. The target mode $\retroreflectivetargetmode$ is an eigenmode on the assumption that it can be retroreflected. However, if there is any clipping loss, the chosen mode cannot be perfectly retroreflected because the field leaking beyond the mirror cannot be retroreflected. This means that we must calculate the true eigenmode of the cavity, $\retroreflectiveeigenmode$, instead of assuming that $\retroreflectivetargetmode$ is an eigenmode. Additionally, the performance calculated for $\retroreflectiveeigenmode$ might not be strictly the maximum possible value, only a lower bound to it, because it was $\retroreflectivetargetmode$, not $\retroreflectiveeigenmode$, that was optimised. In practice, most sensible metrics of cavity performance will discourage clipping loss that significantly reduces the number of cavity round trips. This means that the clipping loss of the target mode $\retroreflectivetargetmode$ is likely to be low, so little power in $\retroreflectivetargetmode$ will leak beyond the mirrors, and therefore $\retroreflectiveeigenmode$ is likely to be very similar to $\retroreflectivetargetmode$. Empirical evidence for this is presented in App.~\ref{app: retroreflective justification}.

Finally, the surface construction method can encounter problems for arbitrary modes with two transverse dimensions, because these can feature phase vortices, which are points where the amplitude goes to zero, but the phase change encountered on a path encircling the vortex does not sum to zero (instead summing to a multiple of $2\pi$). The equiphase surface of a mode with a vortex can require infinitely sharp drops of size $\lambda/2$ in order to maintain phase continuity, often unavoidably through regions of significant mode intensity. While such a mirror surface is a mathematical possibility, it is not a practical solution, and the infinite gradient cliffs would violate the paraxial approximation on which the method rests. To avoid the problem of phase vortices, we restrict our optimisations to situations where the phase only varies with one transverse parameter. For the majority of cases, this parameter will be the cylindrical polar radius $r$ with the assumption of complete azimuthal rotation symmetry, which is guaranteed by using only Laguerre-Gauss modes of $m=0$. We will also use the Hermite-Gauss modes, whose transverse parameter is the Cartesian $x$ coordinate, for cases that break radial symmetry but maintain translational symmetry along the other transverse direction ($y$). Though this is a restriction, we note that many cavities are close to satisfying one of these cases, and for those that are not, one may still obtain qualitative or pedagogical insight.   

To calculate the non-Gaussian eigenmodes of cavities with non-spherical mirrors, we use mode mixing, as described in~\cite{Kleckner:10}, which has been used in a variety of theoretical~\cite{Podoliak:17, Karpov:22_1, Karpov:23, Hughes:23, Hughes:24_1} and experimental studies~\cite{Benedikter:15, Benedikter:19, Walker:21}. This requires the evaluation of the surface through Eq.~(\ref{eq: surface reconstruction}) at multiple transverse coordinates. The modulus is taken to give the smoothest surface (smallest vertical change between adjacent points). In our calculations, if the sum of any two adjacent changes exceeded a phase of $\pi$, the number of samples was deemed insufficient, and the process repeated with double the sampling density.

 \section{Example: Photon extraction from cavities}
 \label{sec: central internal cooperativity}
 \subsection{Identifying the performance metric}
 We begin with the simple example of a single emitter in the centre of an axially symmetric optical cavity, with our goal to maximise the probability of extracting a single photon from the emitter in a single attempt. This figure of merit is very important for emitter-cavity systems used as single photon sources~\cite{Kuhn:10}, but also for systems that produce emitter-photon entanglement to use in quantum repeaters and quantum networks~\cite{Reiserer:15}, where the operation rate depends strongly on the photon emission probability~\cite{Sangouard:09, Krutyanskiy:23b, Monroe:14, Stephenson:20, Krutyanskiy:23a}.
 
 The interaction between the emitter and the cavity mode is traditionally parameterised by $g$, the rate of coherent interaction between the emitter excitation and the cavity mode, $\kappa$, the amplitude decay rate of the cavity mode field, and $\gamma$, the amplitude decay rate of the emitter's excited state to spontaneous emission. The cavity decay rate can be subdivided into two components $\kappa = \kint + \kappa_{T}$, where $\kappa_{T}$ is the transmissive loss through the output mirror, which can generally be controlled, and $\kint$ is the loss to other sources, which is generally minimised as far as possible. Assuming the transmissive loss $\kappa_T$ is freely chosen to maximise the photon extraction probability, it can be shown that the single metric determining the maximum photon extraction probability is
\begin{equation}
\cint = \frac{g^2}{2\kint \gamma},
\label{eq: internal cooperativity rates expression}
\end{equation}
with a maximum extraction probability
\begin{equation}
    \pext = \frac{\sqrt{2\cint +1}-1}{\sqrt{2\cint + 1}+1}
    \label{eq: extraction probability limit}
\end{equation}
where this limit is satisfied in infinite time \cite{Goto:19}. In practice, This bound holds well provided the photon is generated over a time-scale longer than the `critical time' $\tcrit = \max(1/\kappa, \kappa/g^2)$~\cite{Utsugi:22, Hughes:24_2}, which is satisfied in many experimental situations~\cite{Schupp:21}. For a graphical example of how $\cint$ affects $\pext$, see App.~\ref{app: role of internal cooperativity in photon extraction}.

An alternative expression for $\cint$ emphasises the role of the mode profile through the `inverse effective area', $\inverseeffectivearea(\bm{r_e}) = {\epsilon^+}^{2}\left(\bm{r_e}\right)/\int_{P_{z_e}}{\epsilon^+}^2\left(\bm{r}\right) \, d^2\bm{r}$, where $P_{z_e}$ is the transverse plane containing the emitter, which represents the ratio of the intensity of the forward-running cavity mode at the position $\bm{r_e}$ of the emitter to the total forward-running power. The expression is
\begin{equation}
\cint = \frac{3 \lambda^2 \inverseeffectivearea(\bm{r_e})}{\pi \lint},
\label{eq: unit branching internal cooperativity equation}
\end{equation}
where $\lambda$ is the emitter transition wavelength, and $\lint$ is the round-trip power loss to non-transmissive sources (see~\cite{Tanji-suzuki:11, Cox:18} for a derivation, where the internal loss $\lint$, rather than total loss, has been used in order to express $\cint$ rather than the cooperativity $C$, and appropriate quantities have been substituted to match the current notation). Note that the expression assumes, as we shall do throughout, that the emitter is placed perfectly at the antinode of the axial standing wave, which will require a maximum of $\lambda/4$ axial translation of the emitter from its nominal position. It also assumes that the branching ratio of the excited state emission to the cavity mode, which accounts for both the spontaneous emission branching ratio, and the polarisation overlap with the cavity mode, is unity. In the case that the branching ratio is not unity, the right hand side of Eq.~(\ref{eq: unit branching internal cooperativity equation}) should be multiplied by that branching ratio. 

For the case of a cavity with one transverse dimension (specifically, a cavity extended in the $y$-direction so that the $y$-dependence of the cavity mode is negligible), $\cint$ cannot be directly calculated. Instead, we can use the cartesian separability of the mode in this case to assert $\inverseeffectiveareaemitter = \inverseeffectivelengthemitter \inverseeffectivelengthy$, where the inverse effective length in the $x$-direction $\inverseeffectivelengthemitter = {\epsilon^+}^{2}\left(\bm{r_e}\right)/\int_{X_{z_e}}{\epsilon^+}^2\left(\bm{r}\right) \, d\bm{r}$, where $X_{z_e}$ is the $x$-axis in the transverse plane containing the emitter. For a cavity that is infinitely extended in $y$, the inverse effective length in the y direction, $\inverseeffectivelengthy$, is zero, but for any real cavity, $\inverseeffectivelengthy$ will be finite.  We can then optimise the one-transverse dimension cavity using the metric
\begin{equation}
    \cint /\inverseeffectivelengthy = \frac{3 \lambda^2 \inverseeffectivelengthemitter}{\pi \lint}.
\label{eq: single transverse dimension cint}
\end{equation}
This means that $\inverseeffectivelengthy$ never need be calculated to perform optimisation in cases where the $y$-dependence is negligible, but for a real cavity with non-infinite $y$ extent, the value of $\inverseeffectivelengthy$ could be supplied to yield $\cint$.   

 \subsection{Evaluating the performance metric}
 To use the retroreflective optimisation method, we need to be able to calculate $\cintdirect$ or $\cint/ \inverseeffectivelengthy$ directly from the cavity mode, and, in order to use more efficient optimisation algorithms, we can calculate their gradients as well. To do this, we can write the inverse effective area (length) at the emitter $\inverseeffectiveareaemitter$ ($\inverseeffectivelengthemitter$) and internal loss $\lint$ as (unnormalised) matrices. 

The inverse area and inverse length matrix elements are given
\begin{equation}
    \mymatrix{\inverseeffectivearea}_{s', s}(\bm{r}_e) = \left(\epsilon^{LG,+}_{s'}(\bm{r_e})\right)^* \epsilon^{LG,+}_s(\bm{r_e}), \quad \mymatrix{\inverseeffectivelength}_{s', s}(\bm{r}_e) = \left(\epsilon^{HG,+}_{s'}(\bm{r_e})\right)^* \epsilon^{HG,+}_s(\bm{r_e}),
\end{equation} and the loss matrix elements
\begin{equation}
\begin{aligned}
    \lintmatrix &  = \lintmatrix^{+} + \lintmatrix^{-}\\
    \lintmatrix^{\pm} &  = \mymatrix{\mathcal{I}} - \left(1-\frac{1}{2}\lnotclip\right)\left(\mymatrix{\rho}^{\pm}\right)^{\dag}\mymatrix{\rho^{\pm}}, \\
    \mymatrix{\rho}^{\pm}_{s',s} &  = \int_{\Pi_{m_{\pm}}} \left(\epsilon^{\Lambda,+}_{s'}(\bm{r})\right)^* \epsilon^{\Lambda,+}_{s}(\bm{r}) \, d\Pi,
\end{aligned}
\end{equation}
where $\lintmatrix^{\pm}$ is the internal loss matrix of the mirror at positive (negative) $z$ coordinate, $\mymatrix{\mathcal{I}}$ is the identity matrix, $\lnotclip$ is the round trip absorption loss (including scattering), which is the component of the internal round trip loss that does not include clipping, $\mymatrix{\rho}^{\pm}$ is the clipping matrix of the aperture defined by the edge of the mirror at positive (negative) $z$, and $\Pi_{m_{\pm}}$ is the finite domain (a surface for simulations with two transverse dimensions, and a length for a single transverse dimension) of the mirror at positive (negative) $z$. For the spherical apertures considered in the radially symmetric simulations, the elements of $\mymatrix{\rho}^{\pm}$ can be evaluated using the techniques of Vajente~\cite{Vajente:22}. The details of the matrix derivations are given in App.~\ref{app: derivation of matrices for the optimising photon extraction}.

\subsection{Possible improvements in performance}
With the ability to optimise the cavity mode to improve $\cint$, and thus the photon extraction probability, we study the case of an emitter placed in the centre of a cavity of length $L$ with two identical mirrors of diameter $D$. The scenario represents an ideal geometry to couple strongly to the emitter in the centre of the cavity, and features prominently in experimental design~\cite{Takahashi:17_1, Krutyanskiy:23b}. For the purposes of optimisation, we will take the geometric limits $L$ and $D$ as given since in practice $L$ is often restricted by a minimum emitter-mirror distance to mitigate the impact of electric field noise~\cite{Teller:21, Chen:22, Kassa:25}, and the diameter $D$ by the maximum extent of machinable surface or requirements for optical access.

We first consider the case that the mirrors are spherical, and thus the cavity modes are Gaussian. Given a cavity length $L$, there is a minimum viable diameter $\dcrit$, below which every possible Gaussian mode sustains clipping losses that exceed the other internal losses of the cavity $\lnotclip$. In this case, the cavity performance is low. We can derive an approximate expression for $\dcrit$
\begin{equation}
    \dcrit = 2\sqrt{\frac{\lambda L}{\pi}}\sqrt{-\frac{1}{2}\log\left(\lnotclip\right)},
\end{equation}
from the power clipping approximation~\cite{Hunger:10} (see App.~\ref{app: critical diameter and separation}). If the diameter exceeds $\dcrit$, then $\cint$ may be increased by reducing the mirror radius of curvature, and thus focusing the mode more tightly on the emitter, until the increasingly-divergent mode sustains clipping losses on the mirrors that do become significant when compared to $\lnotclip$~\cite{Gao:23}. The spherical mirror shapes, and consequent Gaussian fundamental mode, mean that only the divergence of the cavity mode can be optimised (which determines the waist of the central focus and the waist on the mirrors); there is no flexibility to further control the transverse profile of the mode as it always has a Gaussian decay.

However, the mirrors could be shaped to improve internal cooperativity in this geometry, a suggestion made in Karpov et al.~\cite{Karpov:22_1}. There, as here, the surface was constructed to retroreflect a target mode $\retroreflectivetargetmode$, but in our case $\retroreflectivetargetmode$ is optimised, whereas in their case $\retroreflectivetargetmode$ followed an ansatz. Their $\retroreflectivetargetmode$ is parameterised by a number of occupied basis states $N_p$, with the $N_p$ lowest-order $m=0$ Laguerre-Gauss modes given the same amplitude coefficient, and all other basis states zero amplitude. This mode was chosen because it has the maximum intensity at the emitter for a superposition of those basis states. However, this ansatz mode does have increased divergence compared to the fundamental mode of the basis. For the remainder of this section, we will refer to this method of designing the surface as the `ansatz method'.

\begin{figure}[ht!]
\centering\includegraphics[width=15cm]{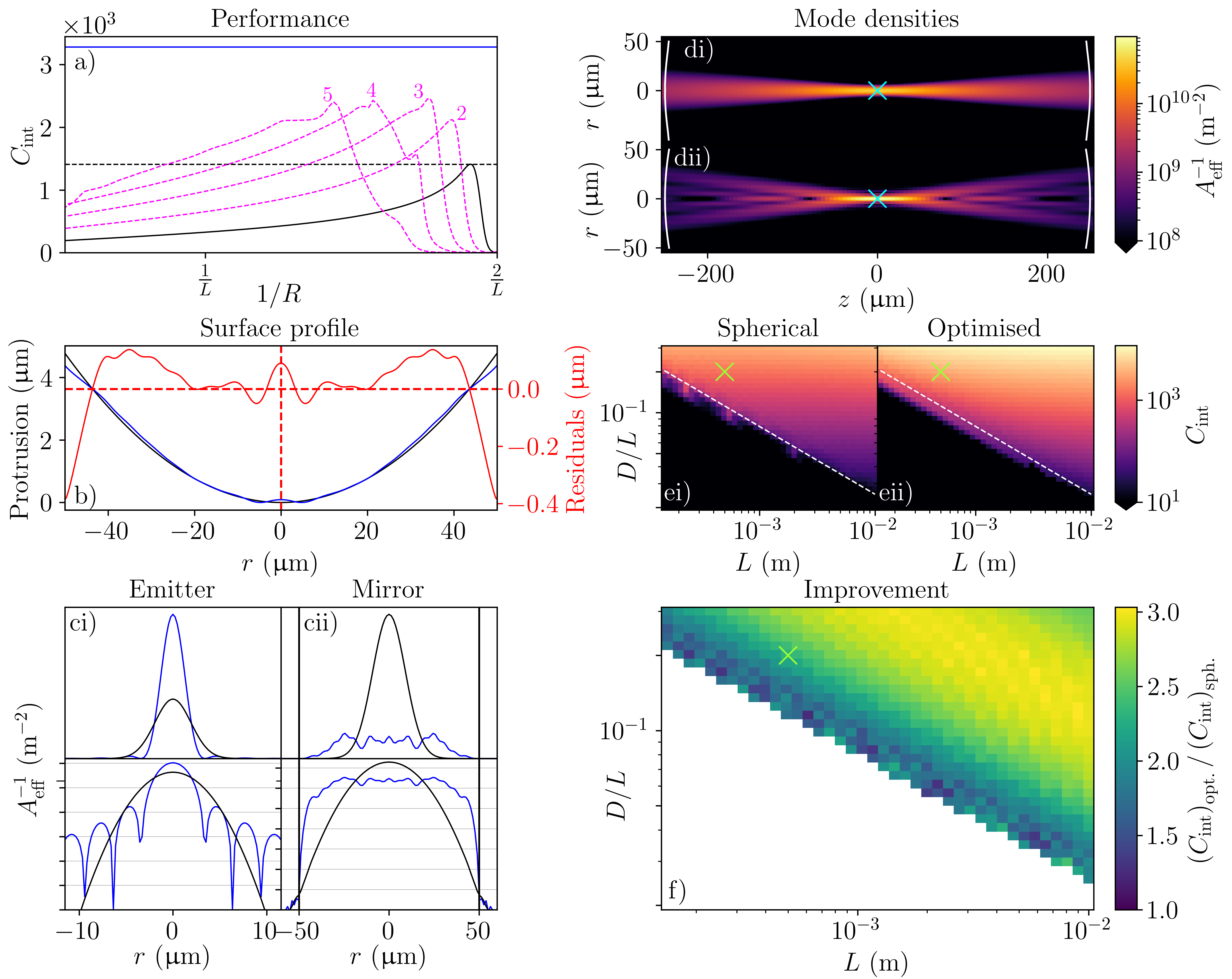}
\caption{Example and summary of improvements in $\cintdirect$ achievable with retroreflective optimisation for a single emitter in the centre of an optical cavity. The example cavity has a length of \SI{500}{\micro\metre}, a mirror diameter of \SI{70}{\micro\metre}, resonant wavelength of \SI{854}{\nano\metre}, and non-clipping internal loss of 20ppm. a) The $\cintdirect$ as a function of nominal mirror radius of curvature $R$ for (black) a spherical mirror and (dashed purple) the mirrors from the ansatz method, where the number of superposed states is labeled around the peak of the corresponding line. Horizontal lines depict the maximum $\cintdirect$ achievable with a (dashed black) spherical mirror, and (blue) retroreflective optimised mirror. b) The surface profile of the best spherical mirror (black) and the optimised mirror (blue), with the residuals overlaid (red). c) Comparison of the mode intensities (linear scale in top row, log-10 scale in bottom row with gridlines indicating decades) for the best spherical mirror cavity (black) and the optimised cavity (blue) in the i) central transverse plane containing the emitter and ii) mirror transverse plane, where the vertical lines mark the mirror edges. d) Mode intensities of the i) best spherical mirror cavity, ii) retroreflective-optimised cavity in an $xz$ cross-section of the mode ($y=0$). The cyan cross indicates the emitter position. e) The maximum $\cintdirect$ achievable with spherical mirrors and optimised mirrors. The white dashed line depicts $\dcrit$. f) The improvement in $\cintdirect$ from choosing the optimised surface over the optimised spherical surface. No data shown where $D<\dcrit$, as these cavities do not have high performance. The green crosses in ei), eii), and f) show the configuration exemplified in panels a)-d).}
\label{fig: central internal cooperativity}
\end{figure}

We investigate the potential of retroreflective optimisation to improve the internal cooperativity compared to any cavity with spherical mirrors, and compared to cavities designed through the ansatz method. In Fig.~\ref{fig: central internal cooperativity}, we study an example situation, and perform the optimisation over a range of cavity geometries to explore trends in the optimisation performance. To perform retroreflective optimisation for a given cavity geometry, we first found the best radius of curvature to use for spherical mirrors, which sets the Laguerre-Gauss basis modes of the optimisation. We then constructed the appropriate matrices to calculate $\cintdirect$ (see Sec.~\ref{sec: central internal cooperativity}) and optimised the real parameter vector $\myvector{C}$ to find $\retroreflectivetargetmode$ with an L-BFGS algorithm~\cite{Zhu:97}. We used the first 171 $m=0$ Laguerre-Gauss modes to construct the matrices, with the first 136 of these corresponding to free coefficients of $\myvector{C}$ (the coefficients of the remaining basis states were set to zero). The number of free coefficients was chosen to be less than the basis size to ensure that the clipping loss of the highest occupied basis state of $\retroreflectivetargetmode$ (corresponding to the final free parameter) was correctly calculated by accounting for its overlap with basis states of somewhat larger $l$-index. The basis size and number of free parameters were chosen to ensure convergence of optimised $\cint$ over the range of geometries investigated in Fig.~\ref{fig: central internal cooperativity}. Details of the basis sizes required for convergence and the parameters used for all simulations in this manuscript can be found in App.~\ref{app: numerical parameters for simulations}.

The first conclusion to draw from Fig.~\ref{fig: central internal cooperativity} is that the retroreflective optimised surface outperforms any spherical surface, and any surface optimised through the retroreflective ansatz method (Fig.~\ref{fig: central internal cooperativity}a). This is expected, because we optimise the retroreflected mode, rather than simply taking an ansatz that should have favourable properties. The surface that achieves this optimised performance is broadly spherical, but features a very particular pattern of distortions at the level of 10's to 100's of nanometers (Fig.~\ref{fig: central internal cooperativity}b). 

We can interpret the performance increase by studying the optimised mode $\retroreflectiveeigenmode$ (Fig.~\ref{fig: central internal cooperativity}c), which is determined via mode mixing on the surface derived from the target mode $\retroreflectivetargetmode$, but is, in practice, very similar to $\retroreflectivetargetmode$. The intensity of the mode on the emitter is strongly increased (Fig.~\ref{fig: central internal cooperativity}ci), which is the origin of the increased coupling. This tighter focussing is achieved by using a more divergent mode that exploits the entire mirror surface more efficiently than any Gaussian mode, as exemplified by the super-Gaussian decay in the intensity at the edge of the mirror evident in the log intensity plot (Fig.~\ref{fig: central internal cooperativity}cii) lower). The intensity density plots (Fig.~\ref{fig: central internal cooperativity}d) indicate how this optimised mode focusses and diverges strongly within the cavity. 

Moving beyond the single example, we can perform the same optimisation procedure for a range of geometries, here varying the length and the mirror diameter. In Fig.~\ref{fig: central internal cooperativity}e), we show the performance achievable with spherical mirrors and with optimised mirrors. The most obvious features are the regions of high and low performance, roughly delineated by the white line depicting $\dcrit$, despite its crude derivation that ignores mode-mixing effects. Retroreflective optimisation can subtly move the exact boundary between the regions, but, similar to the case with spherical mirrors, cannot generally find good performance if $D<\dcrit$. However, retroreflective optimisation can significantly increase $\cintdirect$ in the high performance region ($D>\dcrit$). The scale of improvement tends to increase with increasing mirror diameter before saturating at an improvement factor of roughly 3, which is broadly constant across a large range of length scales. This suggests that ideal mirror shaping could increase internal cooperativity by roughly half an order of magnitude in the central emitter scenario provided the mirror is large enough. This increase is considerable, but the optimised mirror shapes are highly complex, and therefore likely not ideal designs for practical manufacture (see Sec.~\ref{sec: surface profile manufacture} for further discussion on this point). 

Finally, we note that optimisation also changes the $g$ and $\kappa$ parameters of the cavity-QED system. Most directly, the more intense focussing of the cavity mode on the emitter increases the emitter-cavity coupling rate $g$. The optimised cavity $\kint$ typically remains close to the minimum limit set by $\lnotclip$, but the increased $\cint$ means the optimum transmission of the outcoupling mirror will increase, increasing $\kappa$~\cite{Goto:19}. While the impacts of changes to $g$ and $\kappa$ on the photon extraction probability are encapsulated by Eqs.~(\ref{eq: internal cooperativity rates expression}) and (\ref{eq: extraction probability limit}), these rates are also important for achieving strong coupling ($g>\kappa,\gamma$), which is important for observing vacuum-Rabi oscillations, as well as affecting the resonant selectivity of the cavity~\cite{Barrett:18}, and its sensitivity to birefringence~\cite{Kassa:23}. If the experimental application were to place constraints on the acceptable values of $g$ and $\kappa$ beyond a desire to maximise $\pext$, these constraints could be included in the metric optimised to select $\retroreflectivetargetmode$.

\subsection{Non axially-centred emitter}
The case where the emitter is in the centre of the cavity is the most natural for achieving strong coupling, but wider experimental factors may impose dissimilar geometric constraints on the mirrors, or place the emitter away from the centre of the cavity. In this case, we expect the optimised cavity to have an asymmetric mode, with the mirrors on the left and right having different profiles. Retroreflective optimisation can also be used in this case

\begin{figure}[ht!]
\centering\includegraphics[width=13cm]{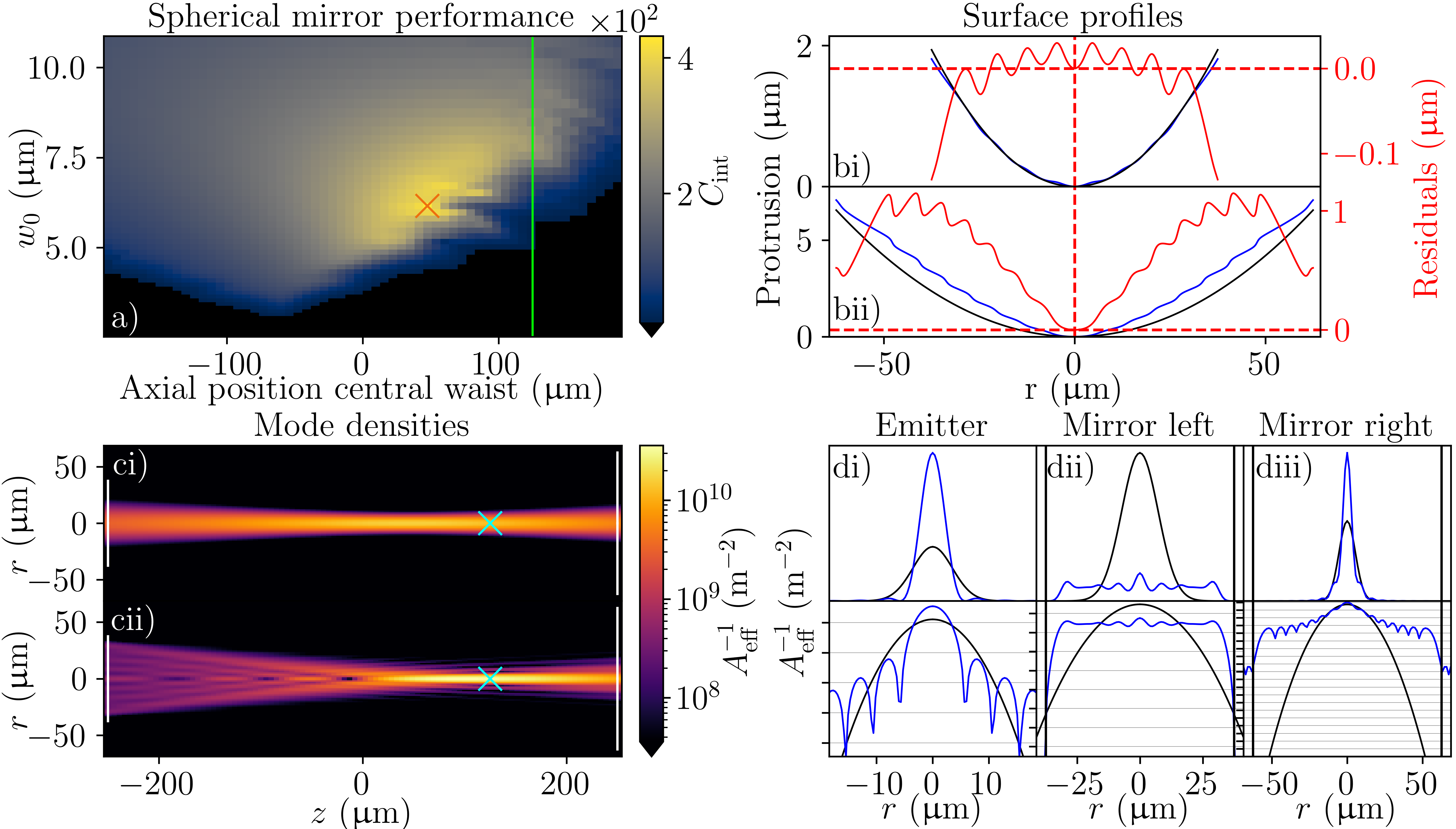}
\caption{Example of improvements in $\cintdirect$ achievable with retroreflective optimisation for a single emitter displaced from the centre of an asymmetric cavity geometry along the cavity axis. The example cavity has a length of \SI{500}{\micro\metre}, left and right mirror diameters of \SI{75}{\micro\metre} and \SI{125}{\micro\metre} respectively, resonant wavelength of \SI{854}{\nano\metre}, non-clipping internal loss of 20ppm, and an emitter placed \SI{125}{\micro\metre} to the right of the cavity centre. a) The $\cintdirect$ for a cavity with spherical mirrors, where the curvatures of the two mirrors are set implicitly by the central waist size and position of the corresponding Gaussian beam family. The green vertical line shows the position of the emitter, and the orange cross the spherical mirror configuration with the highest $\cintdirect$. b) The surface profile of the best spherical mirror (black) and the optimised mirror (blue), with the residuals overlaid (red) for the i) left and ii) right mirror. c) Mode intensities of the i) best spherical mirror cavity, ii) retroreflective-optimised cavity in an $xz$ cross-section of the mode ($y=0$). The cyan cross indicates the emitter position. d) Comparison of the mode intensities (linear scale in top row, log-10 scale in bottom row with gridlines indicating decades) for the best spherical mirror cavity (black) and the optimised cavity (blue) in the i) central transverse plane containing the emitter ii) transverse plane of the left mirror iii) transverse plane of the right mirror, where the vertical lines in ii) and iii) mark the mirror edges.}
\label{fig: non axially centred emitter}
\end{figure}

Example data is shown in Fig.~\ref{fig: non axially centred emitter}, which studies a cavity where the emitter is moved towards the right mirror, and additionally where the right mirror is larger than the left. To achieve the strongest focus at the emitter, the cavity with spherical mirror must find the best compromise between two factors. On one hand, the Gaussian mode would ideally focus on the emitter so that the highest intensity of the mode is at the pertinent location. However, in this case, it is the left mirror, which is both smaller and further from the emitter than the right mirror, that limits the mode divergence. By moving the focus towards the left mirror, more strongly focussing modes can fit inside the cavity without suffering overwhelming clipping loss. Ultimately, this balance means that the best spherical mirrors produce a Gaussian mode that focuses slightly closer to the centre of the cavity than the emitter (Fig.~\ref{fig: non axially centred emitter}a). The retroreflective optimised mode again fits precisely onto the left mirror, whereas the optimised intensity behaves less dramatically on the right mirror, because clipping here does not limit performance (Fig.~\ref{fig: non axially centred emitter}c and d). The optimised mode is also slightly asymmetric about the basis focal position, something which the Gaussian mode cannot do, although the factor 2.75 improvement in $\cint$ is broadly in line with those found in the symmetric case.

\subsection{Transversely-displaced emitter}
Having seen that mirror shaping can improve the coupling cooperativity of a single emitter in the centre of a symmetric cavity, and that improvements can also be found when the cavity geometry is not symmetric about the axis, we now discuss whether the cooperativity can be improved if the emitter is displaced from the centre of the cavity in a transverse direction. As this breaks the azimuthal symmetry, we consider a cavity with just one transverse dimension, using the Hermite-Gauss modes as our basis states.

\begin{figure}[ht!]
\centering\includegraphics[width=16cm]{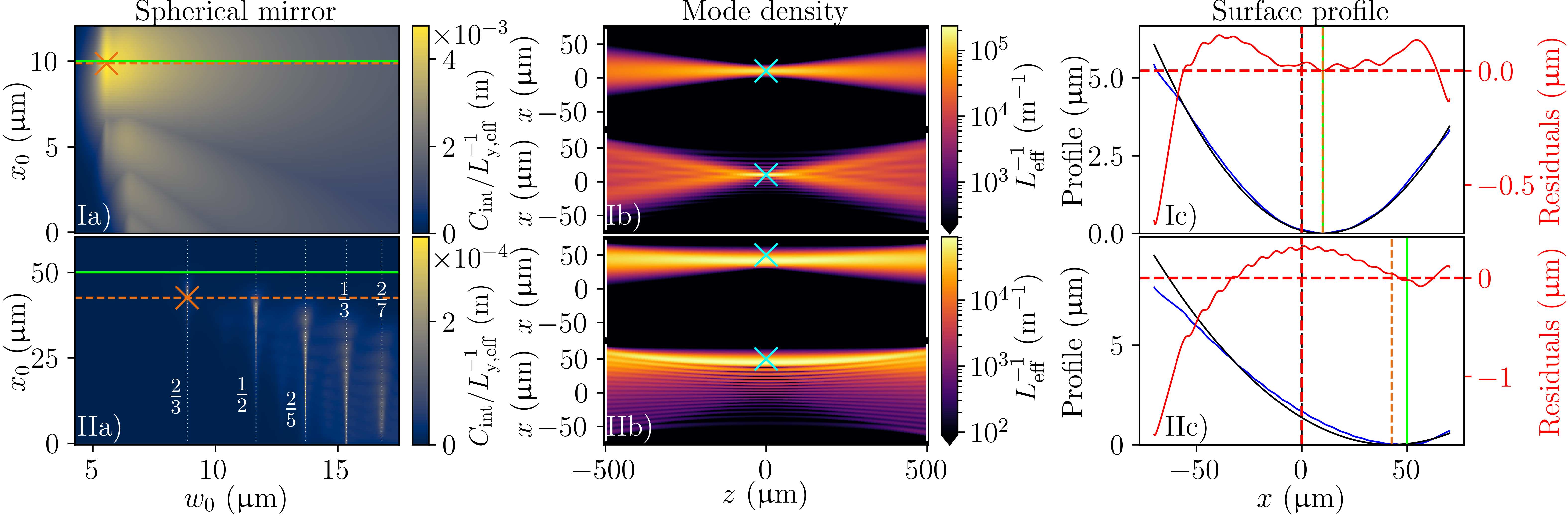}
\caption{Example of improvements in $\cint /\inverseeffectivelengthy$ achievable with retroreflective optimisation for a single emitter displaced from the centre of a cavity geometry in a transverse direction, where the other direction is assumed large. The example cavity has a length of \SI{1}{\milli\metre}, mirror diameter (full mirror length) of \SI{140}{\micro\metre}, resonant wavelength of \SI{854}{\nano\metre}, non-clipping internal loss of 20ppm, and where the emitter is displaced from the centre of the cavity in the transverse direction by row I) \SI{10}{\micro\metre} row II) \SI{50}{\micro\metre}. a) The $\cint/ \inverseeffectivelengthy$ for a cavity with circular mirrors, where the curvature is set implicitly by the central waist size and position of the corresponding Gaussian beam family. The green horizontal line shows the position of the emitter, with the orange cross indicating the best mirror curvature parameters, and orange dotted line the $x$-coordinate of the centre of the best circular mirror. In IIa), the white dotted lines indicate values of $w_{\mathrm{res}}^{q,p}$, where the corresponding fraction labels $q/p$, the transverse mode splitting to free spectral range ratio of the degeneracy. b) Mode intensities (per unit length) of the (top) best circular mirror cavity, (bottom) retroreflective-optimised cavity in the $xz$ cross-section of the mode. The cyan cross indicates the emitter position. c) The surface profile of the best spherical mirror (black) and the optimised mirror (blue), with the residuals overlaid (red). The green solid vertical line indicates the transverse position of the emitter, and the orange dotted line the centre of the best circular mirror.}
\label{fig: transverse displaced emitter}
\end{figure}

We simulate two example cases (Fig.~\ref{fig: transverse displaced emitter}), where the top row shows the emitter displaced a small amount from the axis, and the bottom row the emitter displaced close to in line with the edges of the mirrors. In the former case, we see that the best spherical mirrors focus just slightly more centrally in the cavity than the position of the emitter in order to focus most tightly (Fig.~\ref{fig: transverse displaced emitter}Ia). The optimised mode again uses the space on the mirror more efficiently to achieve a tighter focus (1.65 times improvement in $\cint/\inverseeffectivelengthy$, see Fig.~\ref{fig: transverse displaced emitter}Ib), requiring small but precise modifications to the surface profile (Fig.~\ref{fig: transverse displaced emitter}Ic). 

In the case of large displacement, the physics is somewhat different. Most notably, the spherical mirrors that lead to the highest performance are all found for beam family waists corresponding precisely to transverse degeneracies. The waist values can be predicted as 
\begin{equation}
    w_{\mathrm{res}}^{q,p} = \sqrt{\frac{\lambda L}{2 \pi \tan\left(\frac{\pi q}{2 p}\right)}},
\end{equation}
where $q$ is the number of free spectral ranges between degenerate modes, and $p$ is the increase in the transverse mode index $n$ between degenerate modes. These degeneracies allow the cavity modes to mix in order to avoid clipping on the mirror edge, as can be seen visually in Fig.~\ref{fig: transverse displaced emitter}IIb) (see \cite{Hughes:23} for further discussion on transverse mode mixing). The optimised mode uses asymmetry precisely to create a tightly focused mode with strong divergence towards the cavity centre, but little divergence away from it Fig.~\ref{fig: transverse displaced emitter}IIb) (bottom). This improves $\cint/ \inverseeffectivelengthy$ by 2.45 times from the best circular mirror case, which itself had precisely selected mirror curvature to satisfy the mode degeneracy condition.

 \section{Example: Multiple emitters in cavities}
 \label{sec: multiple emitters}
 The case of the single emitter indicates that mirror surface optimisation can find modes that circumvent traditional limitations of Gaussian beams and consequently improve performance. In that case, the improvement was derived from finding a mode that could fit the shape of the mirror more efficiently than a Gaussian mode, and thus use all the available space on the end mirrors to achieve a tight focus. We now discuss applications involving multiple emitters in a single optical cavity, to ascertain whether improvements could be found for these use cases.

The coupling of multiple emitters to a single cavity mode could unlock two key advantages in quantum information applications. The first is providing direct coupling between these emitters, allowing for quantum gates to be performed~\cite{Duan:03, Takahashi:17_2}. This optical mode interaction could be `long range' compared to traditional interaction vectors, such as motional mode coupling for trapped ions~\cite{Molmer:99} and Rydberg interactions for atoms~\cite{Jaksch:00}, and has thus been proposed to couple small registers of qubits to increase the qubit capacity of single quantum processors~\cite{Ramette:22}. The second advantage is multiplexing the input/output to an optical cavity over many emitters, and thus increase the rate at which the cavity network node interacts with the external world~\cite{Li:24, Sinclair:25, Hu:25, Kikura:25}. Owing to the wide variety of schemes utilising direct intracavity interaction~\cite{Duan:05, Welte:18, Grinkemeyer:25} and the different approaches to multiplexing, we do not attempt to optimise for all metrics of all possible schemes. Instead, we use the interpretation of the cooperativity as roughly representing the number of coherent interactions between a single photon and an emitter before an incoherent process occurs. This means that high cooperativity across all emitters is likely desirable. Thus, our optimisation aim for the entirety of this section will be to maximise the smallest internal cooperativity across all emitters present in the cavity, henceforth written as $\mincint$ for brevity.

\subsection{Coupling of two axial emitters}
\label{subsec: two axial emitters}
\begin{figure}[ht!]
\centering\includegraphics[width=11.5cm]{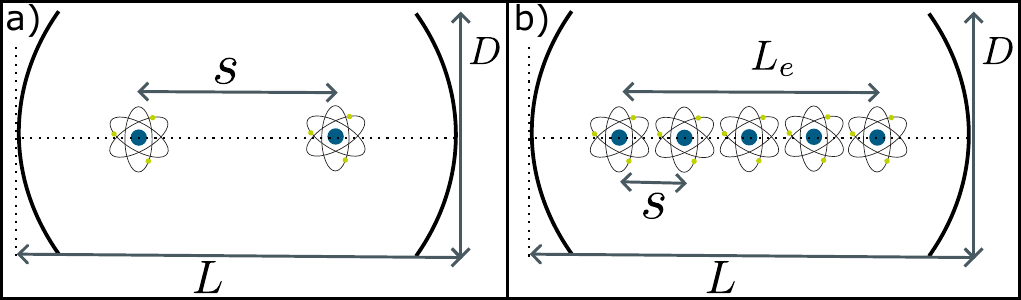}
\caption{Schematic diagrams of the cavity geometries discussed in a) Sec.~\ref{subsec: two axial emitters} and b) Sec.~\ref{subsec: emitters on line}. a) Two emitters separated by $s$ are placed symmetrically on the axis of a cavity of length $L$ with mirror diameter $D$. b) An array of $N_e=5$ emitters of length $L_e$, with the separation between emitters again $s$.}
\label{fig: multi emitter cavity geometries}
\end{figure}

We will start with the simplest case of two identical emitters equally spaced to either side of the centre of the radially symmetric cavity (see Fig.~\ref{fig: multi emitter cavity geometries}a). The enforced axial symmetry of our chosen solutions guarantees that the internal cooperativity is the same on both emitters and therefore we can use the gradient-based optimisation discussed in Sec.~\ref{Sec: mode expansion theory}.  

We first discuss our qualitative expectations of behaviour with spherical mirrors. If the separation $s$ of the emitters is small, the task of focussing strongly on both emitters is identical to the task of focussing to the centre of the cavity. This is the exact case covered in Sec.~\ref{sec: central internal cooperativity}, where the performance is limited by the diameter $D$ of the mirrors, which places an upper limit on the divergence of a low loss cavity mode. However, beyond a certain emitter separation $s$, a cavity mode optimised for central intensity will have diverged significantly at the positions of both emitters, and therefore we would prefer to reduce the mode divergence down from the upper limit. In this regime, the performance is limited by the separation of the emitters, not the diameter of the mirrors. We expect the transition between these two regimes to occur at a separation $\scrit$ for which both limitations are saturated simultaneously. In this case, the best Gaussian mode for minimising beam size on both emitters has clipping loss equal to $\lnotclip$. This leads to
\begin{equation}
    \scrit = \frac{\lambda L^2}{\pi D^2}\left\{\log(2) - \log\left(\lnotclip\right)\right\},
    \label{eq: scrit main text}
\end{equation}
derived in App.~\ref{app: critical diameter and separation}. Assuming $s>\scrit$, the best Gaussian mode is limited by the requirement that it does not diverge strongly between the centre of the cavity and the emitters. As derived in App.~\ref{app: critical diameter and separation}, the optimum radius of curvature and internal cooperativity are
\begin{equation}
    R = \frac{L}{2}\left(1 + \frac{s^2}{L^2}\right), \quad    \mincint = \frac{6 \lambda}{\pi s \lnotclip}.
    \label{eq: best spherical parameters two emitters}
\end{equation}

\begin{figure}[ht!]
\centering\includegraphics[width=16cm]{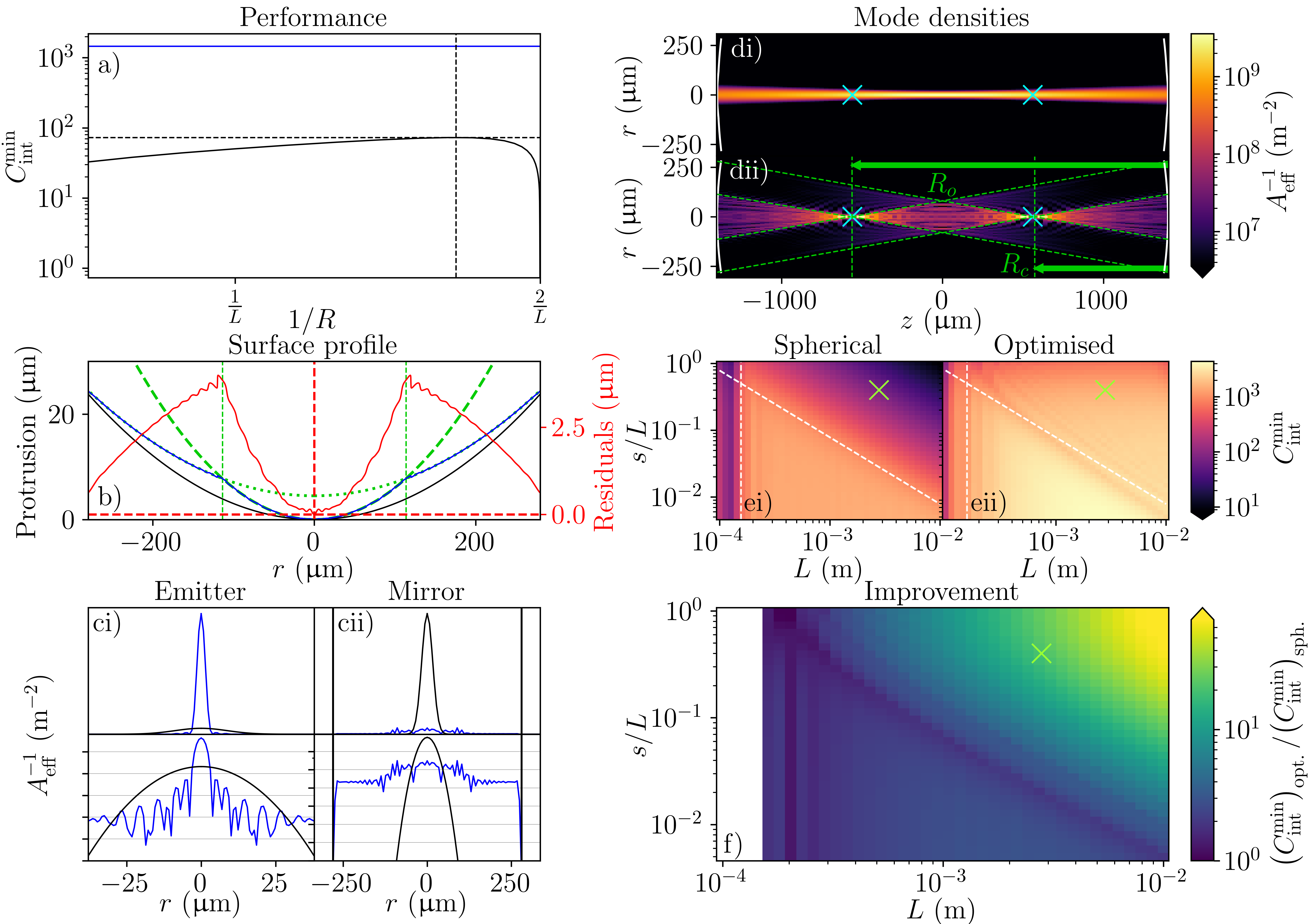}
\caption{Example and summary of improvements in $\mincint$ achievable with retroreflective optimisation for two emitters placed on the axis of the cavity, symmetrically about the central point (see Fig.~\ref{fig: multi emitter cavity geometries}a). The example cavity (panels a-d) has length \SI{2.8}{\milli\metre}, mirror diameter \SI{0.56}{\milli\metre}, resonant wavelength \SI{854}{\nano\metre}, emitter separation \SI{1.12}{\milli\metre}, and non-clipping internal loss 20ppm. a) $\mincint$ across the two emitters for (black line) two identical spherical mirrors as a function of mirror curvature $1/R$ with (dashed black) expected optimum $\mincint$ and $1/R$ from Eq.~(\ref{eq: best spherical parameters two emitters}),  and (horizontal blue line) for retroreflective optimised mirrors. b) Surface profile of the (black) best spherical mirror and (blue) optimised mirror, with (red) residuals overlaid. The optimised surface is fit to a profile with an abrupt transition (dashed green vertical line) between an inner (dotted green line) and outer (dashed green line) spherical profile. c) Mode intensities (linear scale in top row, log-10 scale in bottom row with gridlines indicating decades) for the best spherical mirror cavity (black) and the optimised cavity (blue) in i) a transverse plane containing one of the emitters and ii) the mirror transverse plane, with vertical lines marking the edges of the mirror. d) Mode intensities of the i) best spherical mirror cavity, ii) retroreflective-optimised cavity in an $xz$ cross-section of the mode ($y=0$) with (cyan crosses) the two emitter positions. In ii) The surface fit from b) is indicated through green arrows to show the scale of the central (outer) curvature $R_c$ ($R_o$) and green dashed lines to show the focal points of the central curvature from the transition radius, and the outer curvature from the mirror radius. e) $\mincint$ across the two emitters for spherical mirrors and optimised mirrors. The white dashed diagonal line depicts $\scrit$, and the vertical line delineates $D<\dcrit$ (left) from $D>\dcrit$ (right). f) Improvement in the $\mincint$ value from retroreflective optimisation. No data shown where $D<\dcrit$. The green crosses in ei), eii), and f) show the configuration used for panels a)-d).}
\label{fig: axially separated emitters performance}
\end{figure}

We again study the ability of retroreflective optimisation to improve the performance in this scenario by performing example calculations, as displayed in Fig.~\ref{fig: axially separated emitters performance}. We again select 136 free optimisation parameters in a larger calculation basis of 171 Laguerre-Gauss basis states to converge over the range of geometries. We choose an example for which $s>\scrit$, and find that retroreflective optimisation does significantly increase performance (Fig.~\ref{fig: axially separated emitters performance}a). The optimised mirror shape has an inner and outer radius of curvature, with an abrupt transition between, and interference patterns around that transition (Fig.~\ref{fig: axially separated emitters performance}b). The central fit curvature of the left (right) mirror and outer fit curvature of the right (left) mirror have a geometric focus on the left (right) ion. This produces a mode which focusses strongly on both emitters (Fig.~\ref{fig: axially separated emitters performance}d). When scanning different geometries Fig.~\ref{fig: axially separated emitters performance}e), we see that, where $s<\scrit$, the best spherical and optimised $\mincint$ are fairly constant. This is because the limitation to performance is the ability of the geometry to produce a strongly focussing mode, which does not depend on the separation of the emitters, and is roughly determined by the ratio $D/L$, which is constant in the data. However, where $s>\scrit$, the performance of the cavity with spherical mirrors is now limited by the separation of the emitters, and so decreases as $s$ increases. The optimised mode retains high $\mincint$, because it can focus effectively on both emitters. This can lead to improvements exceeding an order of magnitude in $\mincint$ over a wide range of geometries (Fig.~\ref{fig: axially separated emitters performance}f).

\subsection{Emitters on an axial line}
\label{subsec: emitters on line}

We now discuss the situation of emitters placed on a line in an axial array, again targeting to maximise $\mincint$. We space $N_e$ emitters equally along a line of total length $L_e$ along the axis of the cavity, with the centre of the array coinciding with the centre of the cavity (see Fig.~\ref{fig: multi emitter cavity geometries}b). 

\begin{figure}[ht!]
\centering\includegraphics[width=15cm]{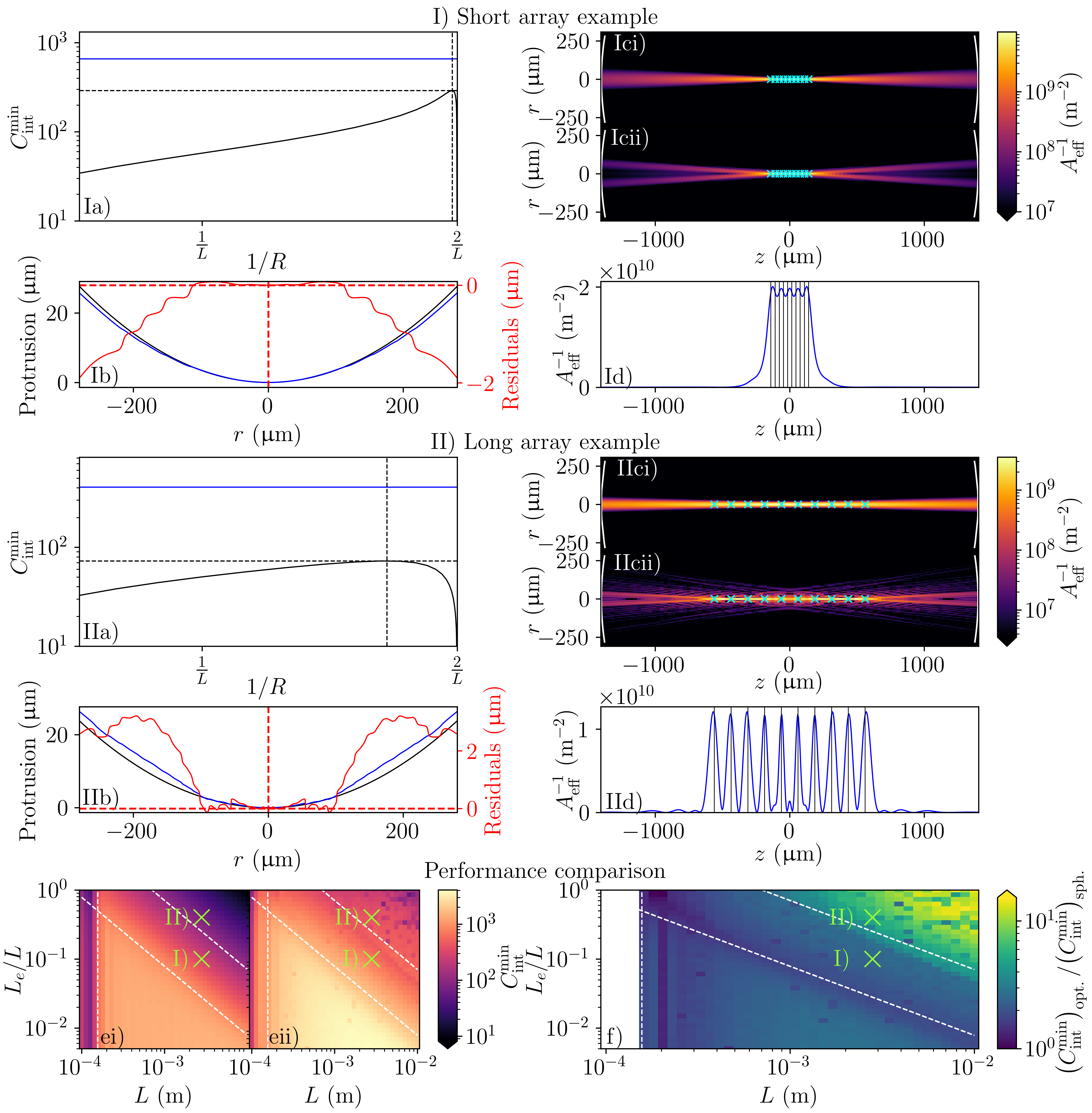}
\caption{Example and summary of improving $\mincint$ across an axial emitter array using retroreflective optimisation. The example cavity has length \SI{2.8}{\milli\metre}, mirror diameter \SI{0.56}{\milli\metre}, resonant wavelength \SI{854}{\nano\metre}, non-clipping internal loss 20ppm, and $N_e=10$ equispaced emitters. The short (long) array of panels I (II) has length \SI{0.28}{\milli\metre} (\SI{1.12}{\milli\metre}). a) Achievable $\mincint$ for (black line) two identical spherical mirrors as a function of mirror curvature and (horizontal blue line) retroreflective optimised mirrors. The dashed black lines show the expected optimum $\mincint$ and corresponding radius of curvature from Eq.~(\ref{eq: best spherical parameters two emitters}), but where the $s$ in that expression is taken be $L_e$, the separation of the two furthest emitters. b) The surface profile of the best spherical mirror (black) and the optimised mirror (blue), with the residuals overlaid (red). c) Mode intensities of the i) best spherical mirror cavity ii) retroreflective-optimised cavity in an $xz$ cross-section of the mode ($y=0$). The cyan crosses mark the emitter positions. d) The intensity of the optimised mode along the cavity axis. The vertical lines indicate the positions of the emitters. e) The maximum $\mincint$ across the array achievable with ei) spherical mirrors and eii) optimised mirrors. The lower white dashed diagonal line depicts $L_e=\scrit$, the upper white diagonal line $s=\scrit$, and the white vertical line $D=\dcrit$. f) The improvement in $\mincint$ from choosing the optimised surface. No data shown where $D<\dcrit$. The green crosses in ei), eii), and f) show the configuration used for panels a)-d), where the cross label `I' or `II' corresponds to the example case.}
\label{fig: emitters on line}
\end{figure}

Example data is shown in Fig.~\ref{fig: emitters on line}, where we optimise to couple the cavity mode to an array of $N_e=10$ emitters. Here, we see three distinct regimes in terms of performance level and improvement, determined by the relation of the length of the array ($L_e$), and the separation of the emitters ($s=L_e/(N_e-1)$) to the critical separation $\scrit$ (see Eq.~(\ref{eq: scrit main text})). In the short array case ($L_e<\scrit$) the array is so short that the optimum fundamental mode is found by focusing the Gaussian beam as tightly as possible to the centre of the cavity. In this case, the spherical mirror and optimised mirror $\mincint$ values are, again, roughly independent of the array geometry, with improvement factors around the level found for the case of Sec.~\ref{sec: central internal cooperativity}. In the second `intermediate' regime $s<\scrit<L_e$, the array length necessitates that the optimum spherical mirror is less concentric than the short array regime, but because $s>\scrit$, the optimised mode cannot focus on each emitter individually without diverging too strongly to be contained by the mirrors of diameter $D$. Empirically, we see that the $\mincint$ of the spherical and optimised mirror cavities drops with increasing array length in this regime. In the final regime, $s<\scrit$, and the optimised mirrors are now able to focus the mode on each emitter individually (Fig.~\ref{fig: emitters on line}IId). This intensity pattern does look like a standing wave, but it actually arises from interference of transverse modes with a periodicity much bigger than the standing wave. In this regime, the $\mincint$ of the spherical mirror cavity continues to fall with increasing array length, but the optimised performance plateaus, allowing for larger improvement factors. This metric could not be optimised with a gradient-based optimiser, so we instead used a simplex optimisation strategy, which is likely the reason for the increased numerical noise on the data. 

These results suggest two distinct situations in terms of coupling to an array of atoms. If we wish to couple to atoms in specific positions, retroreflective optimisation is able to maintain a high $\mincint$ even for long arrays where the performance with spherical mirrors is low. This is achieved by finding modes which focus on each emitter individually. However, if the atoms are not placed in specific positions, but rather anywhere along the axis, we would effectively like to couple strongly everywhere along the array (i.e. $s \rightarrow 0$ with $L_e$ constant). In this case, we are restricted to operating in the intermediate regime and the potential cooperativity improvements from mirror shaping, though significant, are not orders of magnitude. 

\section{Prospect for manufacture}
\label{sec: surface profile manufacture}
The retroreflective optimisation method approximately finds the best possible cavity performance under geometrical constraints for arbitrary mirror shaping. This method is thus well-suited to finding the performance limits imposed by the laws of physics and the corresponding mirror shapes. Perhaps most valuably, it shows how significantly mirror shaping could improve performance in the target application without assuming a limited set of surface parameterisations.

\begin{figure}[ht!]
\centering\includegraphics[width=15cm]{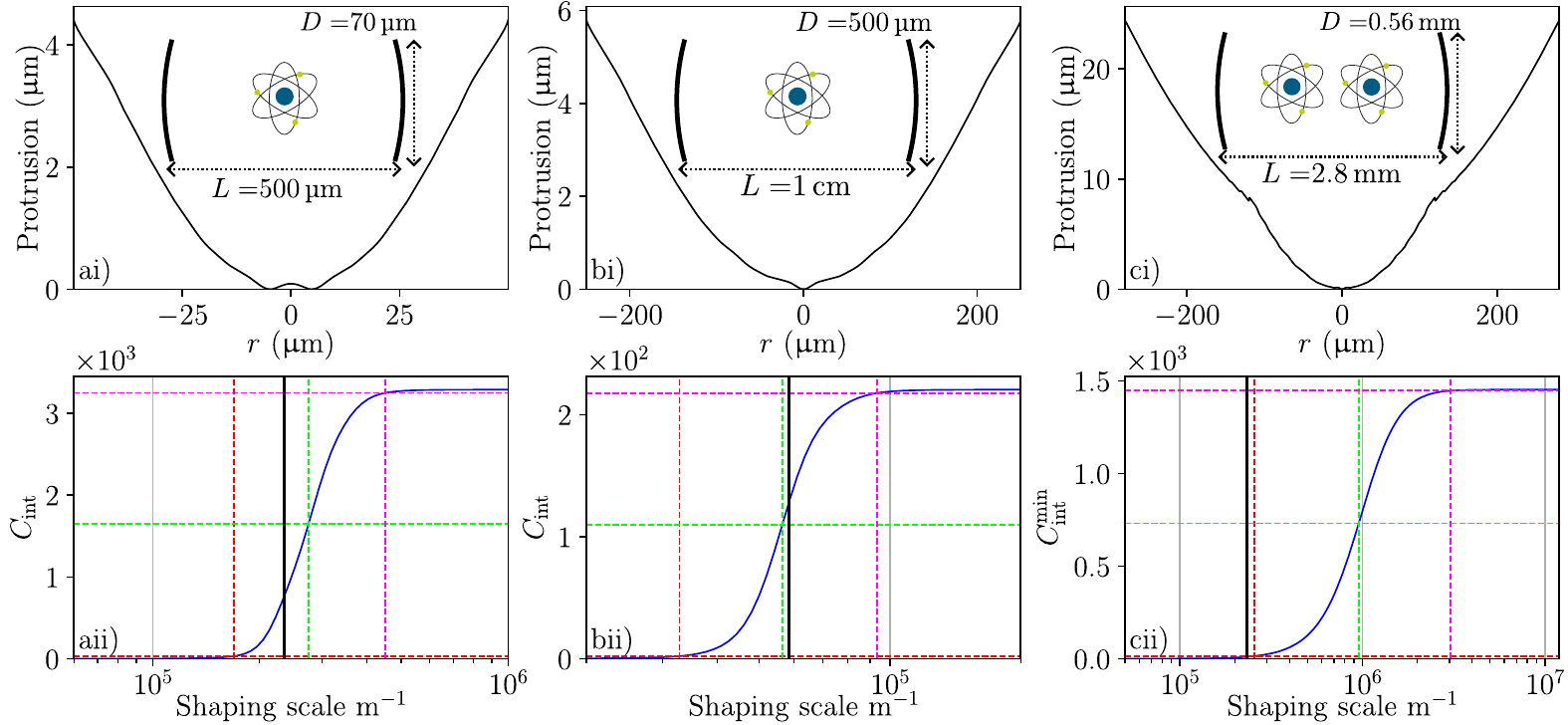}
\caption{The scale of surface shaping required to realise different performance levels for retroreflective optimised mirrors. Row i) The surface profiles of the optimised mirrors, with schematic of the cavity geometry and emitter configuration. Row ii) The performance of the cavity when the mirrors are shaped by spatial scale components up to the specified cutoff spatial frequency. The red, green, and purple crosshairs indicate the shaping precision required to reach 1\%, 50\%, and 99\% of the optimised performance resepctively. The black vertical line indicates $\shapingscale$. a) Example for a cavity coupling to a single central emitter used in Fig.~\ref{fig: central internal cooperativity} ($L=$ \SI{500}{\micro\metre}, $D=$ \SI{70}{\micro\metre}, $\lambda=$ \SI{854}{\nano\metre}, and $\lnotclip=$ 20ppm). b) Example of a different cavity geometry ($L=$ \SI{1}{\centi\metre}, $D=$ \SI{500}{\micro\metre}, $\lambda=$ \SI{854}{\nano\metre}, and $\lnotclip=$ 20ppm) again coupling to a single central emitter. c) Example case from Fig.~\ref{fig: axially separated emitters performance} of a cavity coupling to two emitters placed symmetrically about the centre of the cavity on axis ($L=$ \SI{2.8}{\milli\metre}, $D=$ \SI{0.56}{\milli\metre}, $\lambda=$ \SI{854}{\nano\metre}, and $\lnotclip=$ 20ppm).}
\label{fig: spatial scale}
\end{figure}

The method does not, however, guarantee that the optimised surfaces are easy to manufacture, and the optimised surfaces (see Figs.~\ref{fig: central internal cooperativity}, \ref{fig: non axially centred emitter}, \ref{fig: transverse displaced emitter}, \ref{fig: axially separated emitters performance}, and \ref{fig: emitters on line}) typically harbour high-spatial frequency features. To find the precision of the spatial shaping required to make these mirrors, we take example optimised surfaces, remove frequency components above a cutoff shaping scale with a fourth-order Butterworth filter, and find the performance of cavities with the resulting surfaces. Example results are shown in Fig.~\ref{fig: spatial scale} which indicate (row ii) very low performance if the cutoff scale is too low and near-optimised performance for high spatial resolution. Across the examples, the transition between these low and high performance regions occurs at a shaping scale of approximately
\begin{equation}
    \shapingscale = \frac{D}{\lambda L},
\end{equation}
which is the spatial frequency that would diffract a beam from parallel to the cavity axis to parallel with the line joining the edges of opposite mirrors (i.e. the diffraction allowed by the cavity solid angle). However, the sharp features of the mirror that couples the cavity mode to two emitters (Fig.~\ref{fig: spatial scale} column c) requires shaping slightly more precise than this.

These required shaping scales, though still much larger than the smallest beam diameters of focused ion beam milling~\cite{Trichet:15}, are very demanding. We would therefore recommend that retroreflective optimisation is used as part of a wider method to find promising mirror shapes for manufacture, rather than designing the surfaces itself. In this approach, the retroreflective optimisation method identifies applications for which mirror shaping holds promise to significantly outperform cavities with spherical mirrors, inspires simpler mirror designs that slightly compromise performance for improved manufacturability, and contextualises the performance of these simpler designs against the ultimate physical limits of the geometry. For a full example of this procedure applied to plano-concave cavities see Hughes and Horak~\cite{Hughes:25}.

 \section{Conclusion}
 We have developed a method for optimising optical cavities to maximise user-specified metrics. Instead of directly optimising the mirror surface profile, which is vulnerable to finding local optima in a chosen parameterisation, our method directly optimises the cavity mode, before finding the appropriate surface profile to retroreflect that chosen mode. The mode optimisation procedure takes advantage of a typically simple optimisation landscape, even for large parameter spaces, to find solutions that far outperform standard Gaussian modes. The surface construction component typically works well provided the cavity has low clipping loss, which is generally already required of cavities used for quantum technologies.

 We applied this method to optimise cavities to couple one or many emitters to the cavity mode. We showed that an optimised surface can significantly improve coupling to a single emitter, either central in the cavity or displaced axially or transversely, by using the available mirror space more efficiently. For cases involving multiple emitters along the axis of a cavity, our method found modes with the ability to focus modes on multiple emitters simultaneously, increasing the performance in certain geometric regimes by more than an order of magnitude.

 We anticipate first that our method and results will highlight the general potential of mirror shaping to overcome the limitation imposed by the Gaussian eigenmodes of cavities with spherical mirrors. Secondly, the particularly significant improvements for coupling to multiple emitters, suggest that shaped mirrors could be especially useful for these cases, which find applications in scaling quantum technologies. Finally, though the method tends to produce complex surface profiles that would be challenging to fabricate, these can serve to inspire more manageable designs, and the potential for improvement revealed by retroreflective optimisation could identify applications for which mirror shaping holds promise.

Finally, our method itself relies only upon the paraxial approximation and the assumption of low clipping loss, common operating conditions for Fabry-Perot cavities. This offers considerable flexibility over which metric is optimised, and could allow for the optimisation of cavity geometries in alternative scenarios to those studied here, such as collective coupling to probe ultrastrong coupling~\cite{Johnson:19} or many-body physics~\cite{Lei:23, Pallmann:24}, using vacuum coupling to control chemical reactions~\cite{Andrew:00, Herrera:16, Thomas:19}, or even applications outside cavity quantum electrodynamics.

\begin{acknowledgments}
This work was funded by the UK Engineering and Physical Sciences Research Council Hub in Quantum Computing and Simulation (EP/T001062/1) and Hub for Quantum Computing via Integrated and Interconnected Implementations (EP/Z53318X/1). The authors acknowledge the use of the IRIDIS High Performance Computing Facility, and associated support services at the University of Southampton, in the completion of this work. Data underlying the results presented in this paper are available at Ref.~\cite{dataset}. The authors declare no conflicts of interest.
\end{acknowledgments}

\appendix

\section{Role of internal cooperativity in photon extraction}
\label{app: role of internal cooperativity in photon extraction}
\begin{figure}[ht!]
\centering\includegraphics[width=8cm]{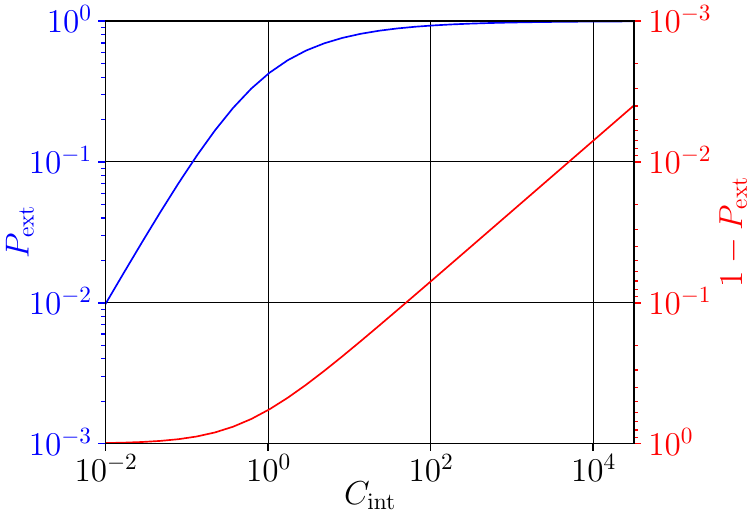}
\caption{The adiabatic upper limit to the probability that a photon is extracted (blue), and the lower limit to the probability that a photon is not extracted (red) from an emitter-cavity system with internal cooperativity $\cint$ whose output transmission is optimised for photon extraction probability at each datapoint.}
\label{fig: internal cooperativity}
\end{figure}

Though the retroreflective optimisation method described in the main text can optimise cavity modes for arbitrary figures of merit, the main text focuses on optimisation of the internal cooperativity due to its critical role in many protocols. Specifically for single photon generation over adiabatic timescales, $\cint$ is the single figure of merit that limits the extraction probability 
\begin{equation}
    \pext = \frac{\sqrt{2\cint +1}-1}{\sqrt{2\cint + 1}+1},
\end{equation}
originally written as Eq.~(\ref{eq: extraction probability limit}) in the main text~\cite{Goto:19}. This extraction probability is visualised in Fig.~\ref{fig: internal cooperativity} so that the practical impact of optimising the internal cooperativity can be understood. For internal cooperativities below 10, increases in the internal cooperativity result in strong increases in $\pext$ (Fig.~\ref{fig: internal cooperativity} blue line). For internal cooperativities above 10, $\pext$ is already a significant fraction of unity, but improvements will dramatically reduce the probability that a photon is not generated (Fig.~\ref{fig: internal cooperativity} red line). The benefits to improvements in the internal cooperativity thus depend upon the $\cint$ of the system before optimisation. It should, however, be noted that the value for $\cint$ quoted in this manuscript (Eq.~(\ref{eq: unit branching internal cooperativity equation})) assumes unit branching ratio to the final state. A real system with non-unity branching ratio will have $\cint$ reduced by a factor of the branching ratio. 

\section{Gradient-based retroreflective optimisation}
\label{app: gradient based retroreflective optimisation}

The retroreflective optimisation method is a two-step process that first optimises the target mode $\retroreflectivetargetmode$ by adjusting the coefficients of its corresponding basis vector $\myvector{C}$, and then constructs the surface to retroreflect this mode. The first stage (the optimisation) can proceed through many different optimisation techniques. A convenient and common case is when the metric to optimise $m$ can be written as the product and quotient of expectation values of matrices with the mode coefficient vector. In this case, it is possible to directly calculate the gradient of $m$ in the parameter space, and thus use optimisation techniques that exploit the direct evaluation of the gradient. In this appendix, we discuss the mathematics of this approach. We begin with a general form of a suitable metric

\begin{equation}
    m = p\left[\frac{\Pi_{i=1}^{i_M} \myvector{C}^T \mymatrix{N}^{(i)} \myvector{C}}{\Pi_{j=1}^{j_M} \myvector{C}^T \mymatrix{D}^{(j)} \myvector{C}}\right]
    \frac{\left(\myvector{C}^T \myvector{C}\right)^{j_M}}{\left(\myvector{C}^T \myvector{C}\right)^{i_M}},
\end{equation}
where $\mymatrix{N}^{(i)}$ is the $i$-th Hermitian matrix of $i_M$ whose expectation is on the numerator, and $\mymatrix{D}^{(j)}$ is the $j$-th Hermitian matrix of $j_M$ whose expectation is on the denominator, and $p$ is a scalar prefactor. The factor on the right hand side is a normalisation so that the metric is independent of the overall scale of the mode coefficients.

For most of the cases in this manuscript, the geometry is symmetric under $z \rightarrow -z$, and we therefore assume that $\abs{\retroreflectivetargetmode}$ is unchanged upon the transformation $z \rightarrow -z$. In this case, the coefficients of $\myvector{C}$ all have the same phase, and thus we choose that all coefficients are real. We will first derive expressions for gradient-based optimisation under this assumption, because it is easy to then adapt those results to cases without this symmetry.

The two derivatives we require are, on the assumption that $\myvector{C}$ is a real vector,
\begin{equation}
    \myvector{\nabla}\left(\myvector{C}^T \myvector{C}\right) = 2 \myvector{C}, \quad \myvector{\nabla} \left(\myvector{C}^T \mymatrix{M} \myvector{C}\right) = 2 \Re(\mymatrix{M} \myvector{C}),
\end{equation}
where $\Re$ signifies the element-wise real part. We can then use the generic formula
\begin{equation}
    \myvector{\nabla}(m) = m \myvector{\nabla}(\ln{m}),
\end{equation}
to yield the gradient
\begin{equation}
    \myvector{\nabla}(m) = 2m \left(\sum_{i=1}^{i_M} \frac{\Re(\mymatrix{N}^{(i)} \myvector{C})}{\expval{\mymatrix{N}^{(i)}}} - \sum_{j=1}^{j_M} \frac{\Re(\mymatrix{D}^{(j)} \myvector{C})}{\expval{\mymatrix{D}^{(j)}}} + (j_M-i_M)\frac{\myvector{C}}{\abs{\myvector{C}}}\right).
    \label{eq: app optimisation gradient}
\end{equation}

Now let us consider the more general case that $\myvector{C}$ has complex coefficients. We can turn this into a real vector $\myvector{C}^{\mathcal{R}}$ with twice the number of coefficients by defining
\begin{equation}
    \myvector{C}^{\Re} = \begin{bmatrix} \Re\left(\myvector{C}\right) \\ \Im\left(\myvector{C}\right)\end{bmatrix},
\end{equation}
where $\Im$ signifies the imaginary part. We can now take any matrix $\mymatrix{A}$ in the basis of $\myvector{C}$ and convert it to a real matrix $\mymatrix{A}^{\Re}$ such that all overlap values are preserved. This means that
\begin{equation}
    \myvector{C}'^{T} \mymatrix{A} \myvector{C} = \left(\myvector{C}'^{\Re}\right)^{T} \mymatrix{A}^{\Re} \myvector{C}^{\Re},
\end{equation}
for all vectors $\myvector{C}$ and $\myvector{C}'$. Equating the elementwise coefficients on this condition yields the method to construct $\mymatrix{A}^{\Re}$ from $\mymatrix{A}$

\begin{equation}
    {\mymatrix{A}^{\Re} = \begin{bmatrix} \mymatrix{A} & i\mymatrix{A} \\ -i\mymatrix{A} & \mymatrix{A} \end{bmatrix}},
\end{equation}

This means that we can now use Eq.~(\ref{eq: app optimisation gradient}) with vectors $\myvector{C}^{\Re}$ and matrices $\mymatrix{A}^{\Re}$ in order to optimise complex vectors (i.e. without restriction to axial reflection symmetry). This form is convenient because many numerical gradient-based optimisation routines naturally work with real vectors.

\section{Justification of the use of retroreflective optimisation}
\label{app: retroreflective justification}
As discussed in Sec.~\ref{sec: retroreflective surface optimisation}, the conceptual validity of retroreflective optimisation requires that the target mode which we optimise is actually an eigenmode of the cavity when we construct the surface. In truth, the cavity eigenmode $\retroreflectiveeigenmode$ is not quite the same as the target mode $\retroreflectivetargetmode$ due to clipping of $\retroreflectivetargetmode$ on the finite diameter mirror. In this case, retroreflective optimisation is not a strict optimisation, but provided that $\retroreflectiveeigenmode$ is similar to $\retroreflectivetargetmode$, it is likely that the retroreflective procedure will find a result that is close to optimal. 

\begin{figure}[ht!]
\centering\includegraphics[width=16cm]{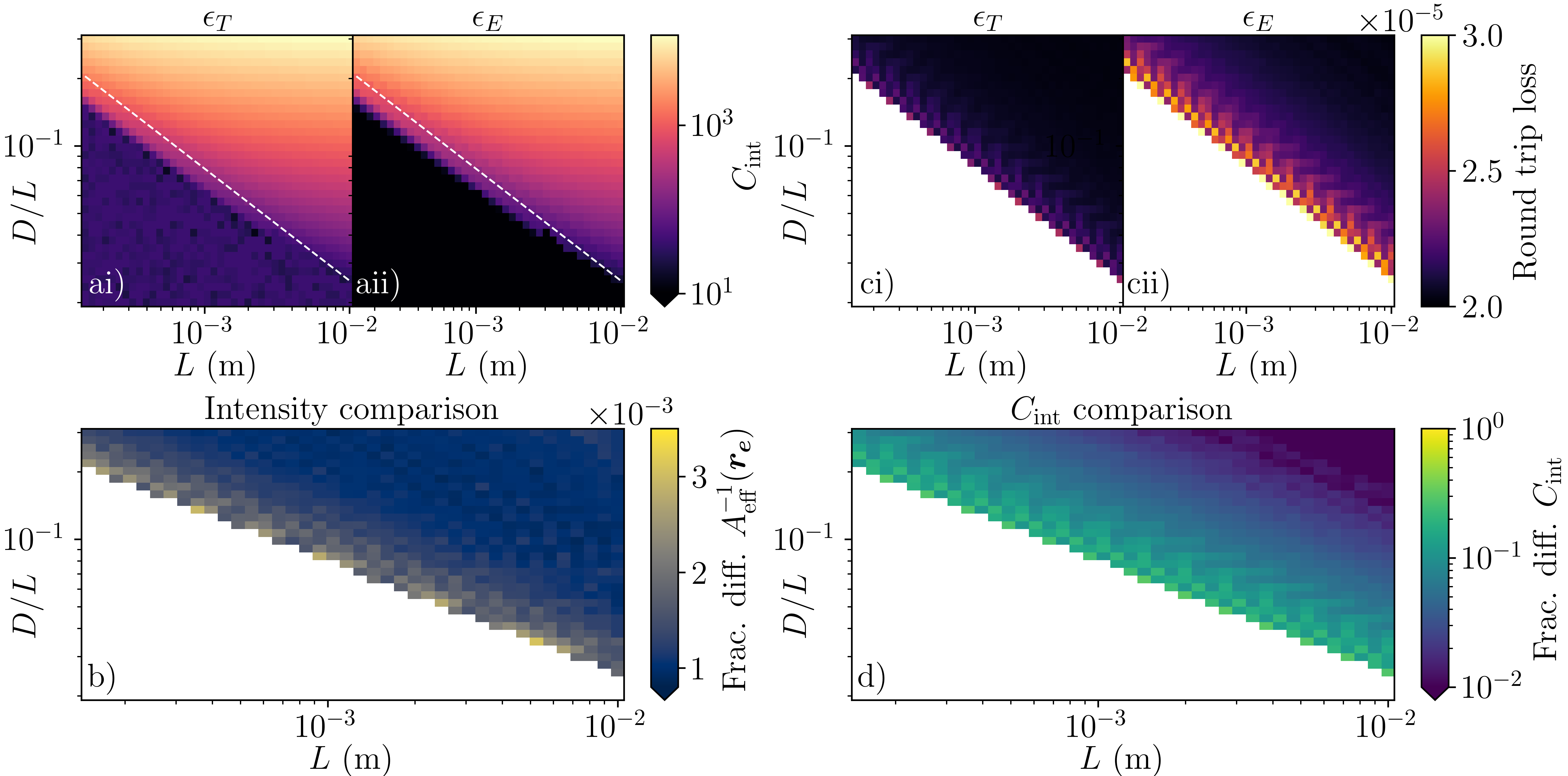}
\caption{Comparison of the target mode $\retroreflectivetargetmode$ and cavity eigenmode $\retroreflectiveeigenmode$ for the dataset used in Fig.~\ref{fig: central internal cooperativity}. a) The $\cintdirect$ of the $\retroreflectivetargetmode$ and $\retroreflectiveeigenmode$. The white dotted line indicates the $\dcrit$. b) The fractional intensity difference at the emitter. c) The round trip loss of the target and eigenmodes. d) The fractional difference in $\cintdirect$. In panels b)-d), data is only shown for cases where the mirror diameter exceeds the critical diameter. The nomenclature `Frac. diff.' in colorbar labels is a shorthand for fractional difference. For a generic quantity $q$ taking value $q_{T}$ for the target mode and $q_{E}$ for the eigenmode, the meaning of `Frac. diff. $q$' is $\abs{q_{T}-q_{E}}/q_{E}$.}
\label{fig: retroreflective justification}
\end{figure}

We check the agreement between $\retroreflectivetargetmode$ and $\retroreflectiveeigenmode$ empirically in Fig.~\ref{fig: retroreflective justification}, which uses the scan data from Fig.~\ref{fig: central internal cooperativity}. We see that (Fig.~\ref{fig: retroreflective justification}a) in the high performance region, where clipping losses are low, the predicted $\cint$ and the eigenmode $\cint$ are very similar, although in the low performance region (where the clipping losses are much higher), the performance values are quite distinct. We can understand this further by comparing the intensity (Fig.~\ref{fig: retroreflective justification}b) and round trip losses (Fig.~\ref{fig: retroreflective justification}c). The intensity comparison shows that $\retroreflectivetargetmode$ and $\retroreflectiveeigenmode$ always have very similar intensities at the emitter, with differences at the part-per-thousand level, showing that $\retroreflectiveeigenmode$ is very similar to $\retroreflectivetargetmode$ in overall spatial structure. However, the loss comparison indicates that the expected loss (calculated from a power clipping method) does not much the true loss (calculated through a mode mixing method). This loss mismatch is most severe when the clipping loss is not negligible, as it is specifically the clipping loss that is not calculated correctly in the retroreflective method. This is not so much of an issue in practice, because one would rarely be interested in cases where the clipping loss is not low, precisely because there would be high clipping loss. This means that the eigenmode $\cint$ matches the expected $\cint$ relatively well in the high-performance region; There are differences on the level of 10-20\% in the reported value, but the optimisation could find improvements on the level of 200\% for this dataset, so it is likely that the retroreflective optimisation procedure does indeed find most of the benefits available from mirror shaping. 

\section{Derivation of matrices for the optimising photon extraction}
\label{app: derivation of matrices for the optimising photon extraction}
As discussed in Sec.~\ref{sec: retroreflective surface optimisation} of the main text, the retroreflective optimisation can proceed more smoothly if quantities within the evaluated metric can be calculated as expectation values of Hermitian matrices. For the case of photon extraction, our chosen metrics are the internal cooperativity $\cintdirect$, and its single-transverse dimension equivalent $\cint/\inverseeffectivelengthy$, which have expressions (taken from Eqs.~(\ref{eq: unit branching internal cooperativity equation}) and (\ref{eq: single transverse dimension cint}) of the main text, but repeated here)
\begin{equation}
\cintdirect = \frac{3 \lambda^2 \inverseeffectivearea(\bm{r_e})}{\pi \lint}, \quad \cint/ \inverseeffectivelengthy = \frac{3 \lambda^2 \inverseeffectivelengthemitter}{\pi \lint},
\label{eq: app unit branching internal cooperativity equation}
\end{equation}
where $\lambda$ is the wavelength, $\lint$ is the fractional round trip loss to non-transmissive sources, $\bm{r_e}$ is the position of the emitter, and $\inverseeffectivearea(\bm{r})$ ($\inverseeffectivelength(\bm{r})$) is the normalised intensity per area (length) of the forward running wave at position $\bm{r}$,
\begin{equation}
    \inverseeffectivearea(\bm{r}) \,\, \mathrm{or} \,\, \inverseeffectivelength(\bm{r}) = \frac{I(\bm{r})}{\int_{\Pi_{z_{r}}}I(\bm{r'})\, d\Pi_{z_{r}}}, 
\end{equation}
where $\Pi_{z_{r}}$ is used to unify the notation for the two cases by representing the infinite transverse domain ($xy$-plane or $x$-axis for two or one transverse dimensions respectively) that passes through $\bm{r}$, and $I(\bm{r})$ is the intensity of the electric field at position $\bm{r}$ (per area for two transverse dimensions, and per length for a single transverse dimension). Of the quantities in Eq.~(\ref{eq: app unit branching internal cooperativity equation}), the inverse effective area $\inverseeffectivearea(\bm{r_e})$, inverse effective length $\inverseeffectivelength(\bm{r_e})$, and internal loss $\lint$ depend upon the cavity mode, and we therefore need to find matrices $\inverseareamatrix(\bm{r_e})$ (or $\inverselengthmatrix(\bm{r_e})$ for single transverse dimension), and $\lintmatrix$ such that
\begin{equation}
    \inverseeffectivearea(\bm{r_e}) = \frac{\myvector{C}^T \inverseareamatrix(\bm{r_e}) \myvector{C}}{\myvector{C}^T \myvector{C}}, \quad \inverseeffectivelength(\bm{r_e}) = \frac{\myvector{C}^T \inverselengthmatrix(\bm{r_e}) \myvector{C}}{\myvector{C}^T \myvector{C}}, \quad \lint = \frac{\myvector{C}^T \lintmatrix \myvector{C}}{\myvector{C}^T \myvector{C}},
\end{equation}
where the normalisation factor is an acknowledgment that the quantities $\inverseeffectivearea(\bm{r})$, $\inverseeffectivelength(\bm{r})$, and $\lint$ are (by design) independent of the total intensity of the mode.  

The first example is the inverse effective area matrix (used for the radially-symmetric situations that comprise the majority of the manuscript). Using the expression for electric field intensity
\begin{equation}
    I(\bm{r}) = \frac{1}{2}c\epsilon_0E(\bm{r})^2,
\end{equation}
where $E(\bm{r})$ is the temporally constant electric field magnitude at position $\bm{r}$, and Eqs.~(\ref{eq: paraxial field definition}) and (\ref{eq: Laguerre Gauss expansion}) from the main text, yields
\begin{equation}
\begin{aligned}
    \inverseareamatrix(\bm{r}) & = \frac{\sum_{s'=0}^{S}\sum_{s=0}^{S} \left(C_{s'}\right)^*\left(\epsilon^{LG,+}_{s'}(\bm{r})\right)^* \epsilon^{LG,+}_{s}(\bm{r}) C_s}{\int_{\Pi_{z_{r}}}\sum_{s'=0}^{S}\sum_{s=0}^{S} \left(C_{s'}\right)^*\left(\epsilon^{LG,+}_{s'}(\bm{r'})\right)^* \epsilon^{LG,+}_{s}(\bm{r'}) C_s \, d^2\bm{r'}}, \\
    &= \frac{\myvector{C}^T \mymatrix{\inverseeffectivearea}(\bm{r}) \myvector{C}}{\myvector{C}^T \myvector{C}}, \\
    \left(\inverseareamatrix(\bm{r})\right)_{s', s} & = \left(\epsilon^{LG,+}_{s'}(\bm{r})\right)^* \epsilon^{LG,+}_{s}(\bm{r}),
\end{aligned}
\end{equation}
where the integrals in the denominator of the first line are simplified using the knowledge that $\{\epsilon_s^{LG, +}\}$ is an orthonormal basis in every transverse plane.

An exactly analogous derivation holds for $\inverselengthmatrix$ on the understanding that, consistent with Eqs.~(\ref{eq: paraxial field definition}) and (\ref{eq: Hermite Gauss mode}) in the main text and the conventions in this appendix, the dimensions of $I$ and $E$ change upon moving to a single transverse dimension. This means that we can use the more compact expression
\begin{equation}
    \left(\inverseareamatrix(\bm{r})\right)_{s', s} \, \, \mathrm{or} \, \, \left(\inverselengthmatrix(\bm{r})\right)_{s', s} = \left(\epsilon^{\Lambda,+}_{s'}(\bm{r})\right)^* \epsilon^{\Lambda,+}_{s}(\bm{r}),
    \label{eq: app inverse area and length}
\end{equation}
on the understanding that $\Lambda = \{LG, HG\}$ is chosen to match the transverse dimensionality of the scenario.

To calculate the internal cooperativity, an evaluation of the internal loss $\lint$ in also required. For this, it is important to clarify the bases of the propagating modes. Conventional mode mixing expresses the mirror at positive (negative) $z$ as a matrix that scatters from the positively (negatively) propagating basis to the negatively (positively) propagating basis. Instead, we want to find our matrices so that they operate purely on the mode coefficients in the positively propagating basis, with the negatively propagating mode completely specified by the assumption of perfect retroreflection (this avoids complications in the Laguerre-Gauss basis where retroreflection swaps coefficients from $m \rightarrow -m$ because retroreflection preserves helicity not angular momentum). With this in mind, it is easiest to start with the mirror at positive $z$. Imagine replacing the mirror with an aperture of the same size, where light passing through the aperture sustains combined scattering and absorption losses $\lnotclip/2$, to match the definition of $\lnotclip$ from the main text as the \textbf{round trip} non-clipping internal loss. The loss upon reflection from the mirror is the same as the loss of power upon this imagined transmission as the only further action of the mirror is to redirect power. The coefficients $\myvector{C}_{\mathrm{out}}$ of the mode superposition representing this imagined transmission can be written in terms of the input coefficients ($\myvector{C}$) with
\begin{equation}
    \left(\myvector{C}_{\mathrm{out}}\right)_{s'} = \sqrt{1 -\frac{1}{2} \lnotclip}\sum_{s=0}^S \mymatrix{\rho}_{s',s} \myvector{C}_s, \quad
    \mymatrix{\rho}^{+}_{s',s} = \int_{\Pi_{m_{+}}} \left(\epsilon^{\mathrm{\Lambda,+}}_{s'}(\bm{r})\right)^* \epsilon^{\mathrm{\Lambda,+}}_{s}(\bm{r}) \, d\Pi,
    \label{eq: app single mirror lint coefficient change}
\end{equation}
where $\mymatrix{\rho}^{+}$ is the clipping matrix of the aperture defined by the edge of the mirror at positive $z$ (which is to what its $+$ superscript refers), and $\Pi_{m_+}$ is the finite domain of the mirror at positive $z$. The internal loss matrix of the mirror can then be derived

\begin{equation}
    \mathcal{L}_{\mathrm{int}}^{+}  = \frac{\left(\myvector{C}\right)^{\dag}\myvector{C} - \left(\myvector{C}_{\mathrm{out}}^{L/R}\right)^{\dag}\myvector{C}_{\mathrm{out}}^{L/R}}{\left(\myvector{C}\right)^{\dag}\myvector{C}}, \quad
    \mymatrix{\mathcal{L}}_{\mathrm{int}}^{+}  =  \mymatrix{\mathcal{I}} - \left(1- \frac{1}{2}\lnotclip\right)\mymatrix{\rho^{+}}^{\dag}\mymatrix{\rho^{+}} ,
\end{equation}

For the mirror at negative $z$, we can repeat this process with the the negatively propagating mode $\epsilon^{-}$ to see that we would have
\begin{equation}
    \left[\mymatrix{\mathcal{L}}_{\mathrm{int}}^{-}\right]_{\mathrm{neg. basis}}  =  \mymatrix{\mathcal{I}} - \left(1- \frac{1}{2}\lnotclip\right)\mymatrix{\rho^{-}}^{\dag}\mymatrix{\rho^{-}} , \quad
    \mymatrix{\rho}^{-}_{s',s}  = \int_{\Pi_{m_{-}}} \left(\epsilon^{\mathrm{\Lambda,-}}_{s'}(\bm{r})\right)^* \epsilon^{\mathrm{\Lambda,-}}_{s}(\bm{r}) \, d\Pi,
\end{equation}
for an expression of $\mymatrix{\mathcal{L}}_{\mathrm{int}}^{-}$ in the negative propagating basis (whatever those basis states happen to be). To convert this to the positive propagating basis, we note that $\mymatrix{\rho}$ is an integral of the normalised intensity matrix (i.e. inverse effective area or inverse effective length) over the surface of the mirror (see Eq.~(\ref{eq: app inverse area and length})). Because reflection does not translate power transversely, within the domain of the mirrors we know that the intensity of the incoming mode and outgoing mode are the same everywhere. Therefore, over the mirror domain we can replace the intensity matrix of $\epsilon^{\Lambda,-}$ with the intensity matrix of $\epsilon^{\Lambda,+}$. Combining these results, we can write the internal loss matrices of both mirrors as

\begin{equation}
    \mymatrix{\mathcal{L}}_{\mathrm{int}}^{\pm}  =  \mymatrix{\mathcal{I}} - \left(1- \frac{1}{2}\lnotclip\right)\mymatrix{\rho^{\pm}}^{\dag}\mymatrix{\rho^{\pm}} , \quad
    \mymatrix{\rho}^{\pm}_{s',s}  = \int_{\Pi_{m_{\pm}}} \left(\epsilon^{\mathrm{\Lambda,+}}_{s'}(\bm{r})\right)^* \epsilon^{\mathrm{\Lambda,+}}_{s}(\bm{r}) \, d\Pi,
\end{equation}
which is now a form that takes mode coefficients purely in the positive propagating basis. In the case that the mirror apertures have identical diameter and distance from the origin, and that the mode is restricted to be even symmetric with respect to $z$-reflection about the origin, the internal loss matrices for the two mirrors are identical. This can help reduce calculation time.

\section{Derivation of critical diameter, separation, and best geometry for coupling to two emitters}
\label{app: critical diameter and separation}

The manuscript uses concepts of a critical diameter $\dcrit$ and critical separation $\scrit$. The critical diameter is the diameter of the mirror for which the clipping losses of the most compact possible Gaussian mode at the mirror equal the non-clipping losses, and is therefore a rough threshold for when a mirror has large enough diameter for the cavity to harbour a mode with negligible clipping loss. The critical separation is used for cases with axially-separated emitters, and it represents the threshold in emitter separation above which the divergence of the mode is limited by the requirement to stay focused on both emitters, rather than being limited by clipping on the mirrors. It turns out that, while deriving $\dcrit$ and $\scrit$, one can easily find the best geometry for coupling to two axial emitters, and therefore the derivation of all three are combined into a single appendix.

Assume that we have restricted the geometry of a concave-concave cavity to have length $L$ and mirror diameter $D$. The identical mirrors both have a radius of curvature $R$, which we assume can be freely set. 

A rigorous calculation of the expected clipping loss of the cavity is complicated, and should account for mode-mixing effects~\cite{Hughes:23}. However, in the interests of obtaining analytical expressions that capture the overarching impact of the geometry, we can approximate the clipping loss through the power clipping method~\cite{Hunger:10}, which takes the mode of standard cavity theory (derived assuming infinite mirrors), and then calculates the fraction of the incident power lying outside the finite mirror on each reflection. For mirrors of diameter $D$, this leads to a \textbf{round trip} clipping loss
\begin{equation}
    \lclip = 2 \exp{-2\left(\frac{D}{2w_m}\right)^2},
    \label{eq: app critical diameter clipping}
\end{equation} 
where $w_m$ is the waist of the mode at the mirror.

It is well known from standard cavity theory~\cite{Siegman:86} that to minimise the size of the beam waist $w_P = w(z_P)$ at point $P$ on the beam axis with axial coordinate $z_P$ requires a cavity and cavity mode with the following properties
\begin{equation}
    z_0 = z_P, \quad R = \frac{L}{2}\left(1 + \left(\frac{2z_P}{L}\right)^2\right), \quad w_0 = \sqrt{\frac{\lambda z_P}{\pi}}, \quad w_P = \sqrt{\frac{2\lambda z_P}{\pi}},
    \label{eq: app minimum waist cavity properties}
\end{equation}
which will be useful for the rest of this appendix.

Starting with the derivation of $\dcrit$, we are concerned with the minimum possible clipping loss for any Gaussian mode, given the constraints of $L$ and $D$, which is obtained when $w_m$ is minimised. We aim to find the critical diameter $D=\dcrit$ such that the clipping loss $\lclip$ and non-clipping loss $\lnotclip$ balance. This is because, if $\lclip$ dominates, the total cavity losses are high, and the performance low for any spherical mirror, whereas if $\lclip$ is negligible, it is possible to achieve loss at the minimum level of $\lnotclip$, and thus achieve high cavity performance. The critical diameter achieving this balance is therefore
\begin{equation}
    \dcrit = \sqrt{\frac{\lambda L}{\pi}}\sqrt{2\left(\log(2)-\log\left(\lnotclip\right)\right)},
\end{equation}
obtained through rearrangement of Eq.~(\ref{eq: app critical diameter clipping}) with $w_m=\sqrt{\lambda L/\pi}$ as the minimum waist at mirror from Eq.~(\ref{eq: app minimum waist cavity properties}).

We now derive the critical separation $\scrit$, defined such that the mode that has the minimum waist size at both emitters separated by the critical separation $\scrit$, must have clipping loss on the mirror $\lnotclip$. We know (Eq.~(\ref{eq: app minimum waist cavity properties})) that the smallest beam waists in the emitter planes (separated by $s$), are found with central waist $w_0 = \sqrt{\lambda s / (2 \pi)}$. This leads to a waist at the mirror of
\begin{equation}
    w_m = \sqrt{\frac{\lambda (L ^ 2 + s ^ 2)}{2\pi s}} \approx \sqrt{\frac{\lambda L ^ 2}{2\pi s}},
\end{equation}
where the approximation that $L^2 \gg s^2$ has been made in the right hand expression for algebraic convenience. This approximation is often reasonable, as the emitter separation could be far smaller than the cavity length, but not bigger. However,  we would expect to see some discrepancies when $s \sim L$. Taking this assumption, substituting $w_m$ into Eq.~(\ref{eq: app critical diameter clipping}) where the clipping loss should equal the non-clipping internal loss $(\lclip=\lnotclip)$, and rearranging, yields
\begin{equation}
    \scrit = \frac{\lambda L^2}{\pi D^2}\left\{\log(2) - \log\left(\lnotclip\right)\right\}.
\end{equation}

Finally, we discuss the large mirror diameter $D$ limit to $\cintdirect$ when coupling to two emitters separated by $s$. As mirror diameter is large, the clipping losses will be negligible, so our only concern is minimising the waist on both emitters. From Eq.~(\ref{eq: app minimum waist cavity properties}), we know the minimum waist at the emitters is $w_e = \sqrt{\lambda s/\pi}$.

The inverse effective area at the emitter for this mode is 
\begin{equation}
    \inverseeffectivearea = \frac{2}{\pi} \frac{1}{w_e^2}.
\end{equation}
Additionally, as the clipping losses are expected to be negligible, we expect $\lint=\lnotclip$. Applying Eq.~(\ref{eq: unit branching internal cooperativity equation}) of the main text results in an internal cooperativity of
\begin{equation}
    \cintdirect = \frac{6 \lambda}{\pi s \lnotclip}.
\end{equation}

\section{Numerical parameters for simulations}
\label{app: numerical parameters for simulations}

The data presented in the main text compares retroreflective optimised cavities against the cavities with the best possible spherical/circular mirrors for the same geometric constraints. This procedure has three steps. The first is to scan the possible spherical/circular mirrors in order to find both the best performance with spherical mirrors, and the basis functions to use for retroreflective optimisation. These calculations were performed using mode mixing using $\basissizespherical$ basis states and numerical integration with $\samplesspherical$ samples in the Riemann sum (linearly spaced in either $r$ or $x$ coordinate). The second step is the first component of the retroreflective optimisation i.e. the determination of $\retroreflectivetargetmode$. This optimises the chosen metric on the assumption of perfect retroreflection, where the chosen metric is evaluated in a basis of size $\basissize$, of which we choose to optimise the coefficients of the first $\basissizeopt$ basis states, and leave the remainder zero. The third step is to take the optimised $\retroreflectivetargetmode$, and use the surface reconstruction method (Sec.~\ref{sec: retroreflective surface optimisation} main text), where the surface is initially sampled (again, linearly in $r$ or $x$) using $\samples$ small elements, and mode mixing is performed in a basis of size $\basissize$. A summary of the parameters used for the simulations can be found in Table~\ref{tab: numerical parameters for simulations}. From this, we highlight that the optimisation for Fig.~\ref{fig: transverse displaced emitter}, where the emitter was displaced towards the edge of the cavity, required a very large number of samples for its mode mixing integrals to converge, but such sampling was not necessary for cases with the emitter closer to the cavity centre.

\begin{table}[h]
    \centering
    
    \begin{tabular}{|c|c|c|c|c|c|}
        \hline
        Example & $\basissizespherical$ & $\samplesspherical$ & $\basissize$ & $\basissizeopt$ & $\samples$ \\
        \hline
        Fig.~\ref{fig: central internal cooperativity} & 171 & 15000 & 171 & 136 & 15000 \\
        \hline
        Fig.~\ref{fig: non axially centred emitter} & 171 & 15000 & 171 & 136 & 15000 \\
        \hline
        Fig.~\ref{fig: transverse displaced emitter} & 50 & 300 & 250 & 200 & 500000 \\
        \hline
        Fig.~\ref{fig: axially separated emitters performance} examples & 171 & 15000 & 171 & 136 & 15000 \\
        \hline
        Fig.~\ref{fig: axially separated emitters performance} scan & 171 & 5000 & 171 & 136 & 5000 \\
        \hline
        Fig.~\ref{fig: emitters on line} examples & 171 & 15000 & 171 & 136 & 15000 \\
        \hline
        Fig.~\ref{fig: emitters on line} scan & 111 & 5000 & 111 & 81 & 5000 \\
        \hline
    \end{tabular}
    \caption{Parameters used for the retroreflective optimisations conducted in the main text. The columns indicate the parameter, and the rows discuss the figure, or part of the figure, where the relevant optimisation is used.}
    \label{tab: numerical parameters for simulations}
\end{table}

\begin{figure}[ht!]
\centering\includegraphics[width=15cm]{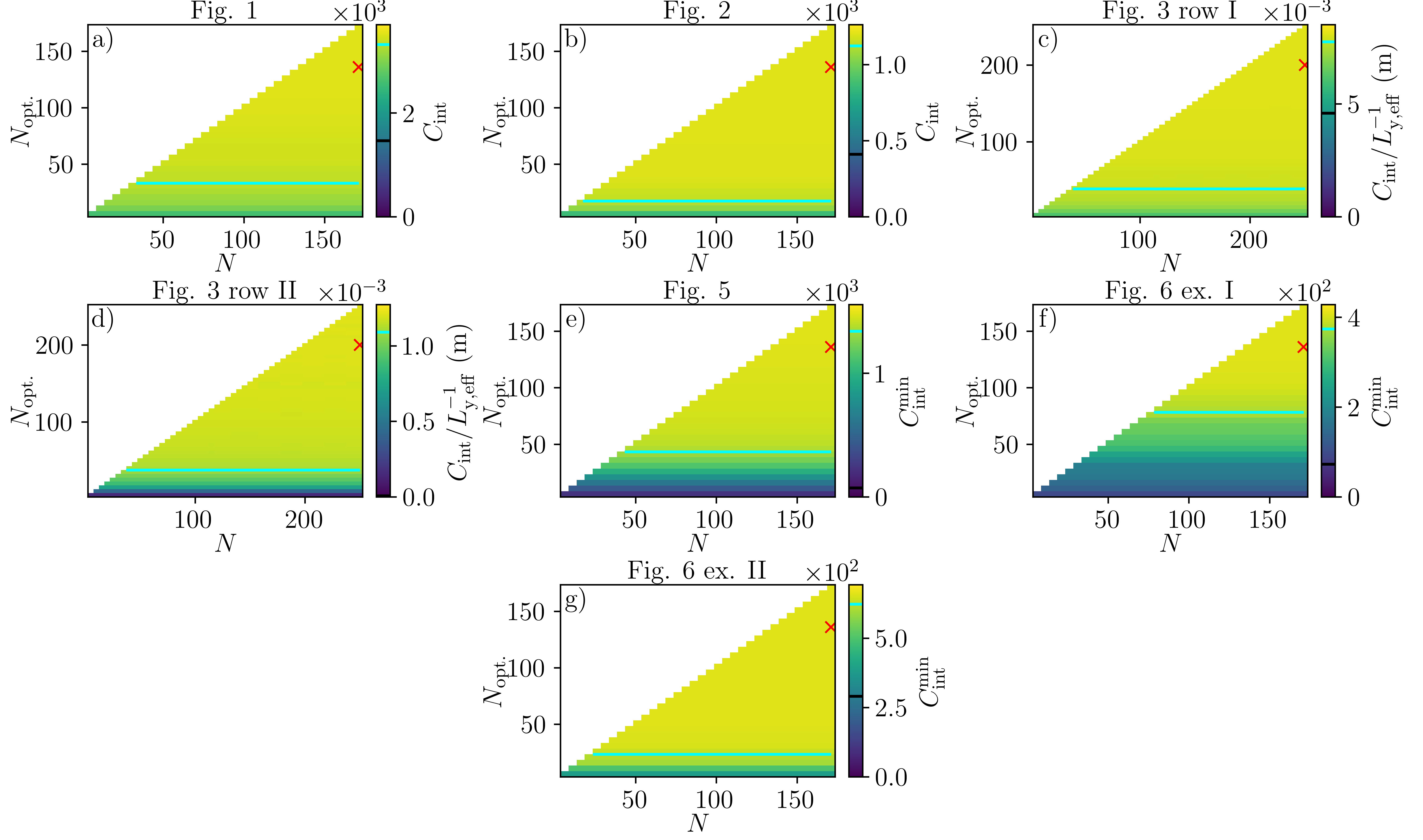}
\caption{The dependence of the cavity performance assuming perfect retroreflection on the basis size $\basissize$ and number of variable coefficients $\basissizeopt$, where $\basissizeopt \leq \basissize$. Note that, because this studies the convergence of the first part of retroreflective optimisation only, no mode mixing calculation is conducted. Each panel shows the convergence data for the titled figure. The black band on the colour bar indicates the retroreflected expected performance for the best spherical mirror. The cyan band in the colour bar and as a contour on the performance map indicates the boundary above which the reported performance has increased by 90\% of the total increase from the spherical case to the optimised value reported for the largest basis sizes in that panel's data. The red cross indicates the parameters chosen for the case study examples in each figure. Note that the black band in panel d) is very close to zero, because the best spherical mirror cavity produced a highly mixed mode, where the nominal fundamental mode was strongly clipped.}
\label{fig: basis convergence}
\end{figure}

The second part of this procedure (i.e. the first part of the retroreflective optimisation method) optimises $\retroreflectivetargetmode$ by changing the coefficients of $\basissizeopt$ states in a basis of size $\basissize$. The number of coefficients ultimately corresponds to the number of degrees of freedom in the target mode, and consequently the resultant mirror surfaces. If $\basissizeopt$ is too low, there will be not enough freedom to describe the optimum $\retroreflectivetargetmode$, and thus the method will not report the optimum performance. Data indicating how the retroreflective performance (i.e. the cavity performance assuming perfect retroreflection) varies with $\basissizeopt$ and $\basissize$ are shown in Fig.~\ref{fig: basis convergence}. Here we see that, for most examples, $\leq 50$ basis states and optimised coefficients are sufficient to find the optimised performance to a high degree of accuracy.

\end{document}